\documentclass{article}

\newcommand{\delt}{1.2}
\newcommand{\scala}{1.2}

\usepackage{amsfonts,latexsym,upref}
\newcommand{\rappr}[1]{
    \raisebox{-0.5 ex}[0 ex][0 ex]
     {\begin{picture}(10,10)(0,0)
    \put(5,5){\circle{10}}
    \put(3,2.5){${#1}$}
    \end{picture}}}
\newcommand{\rapprmedio}[1]{
    \raisebox{-0.5 ex}[0 ex][0 ex]
     {\begin{picture}(14,10)(0,0)
    \put(7,5){\oval(14,10)}
    \put(3,2.5){${#1}$}
    \end{picture}}}
\newcommand{\rapprmediocre}[1]{
    \raisebox{-0.5 ex}[0 ex][0 ex]
     {\begin{picture}(22,10)(0,0)
    \put(11,5){\oval(22,10)}
    \put(3,2.5){${#1}$}
    \end{picture}}}
\newcommand{\rapprlungo}[1]{
    \raisebox{-0.5 ex}[0 ex][0 ex]
     {\begin{picture}(70,10)(0,0)
    \put(35,5){\oval(70,10)}
    \put(3,2.5){${#1}$}
    \end{picture}}}

\newcommand{\lst}{{\it List}}
\newcommand{\bag}{{\it MSet}}

\newcommand{\clist}{{\it CList}}
\newcommand{\set}{{\it Set}}
\newcommand{\Th}{\mathbb{T}}
\newcommand{\T}{{\cal T}}
\newcommand{\size}{\mathit{size}}
\newcommand{\false}{{\tt false}}
\newcommand{\true}{{\tt true}}
\newcommand{\nil}{{\tt nil}}
\newcommand{\e}{\emptyset}
\newcommand{\mo}{\{  \hspace*{-0.55ex} [\,}
\newcommand{\mc}{\,] \hspace*{-0.55ex} \}}
\newcommand{\clo}{[  \hspace*{-0.35ex} [\,}
\newcommand{\clc}{\,] \hspace*{-0.35ex} ]}
\newcommand{\lf}{[\, \cdot \,|\,\cdot \,]}
\newcommand{\lcf}{\clo\,\cdot\,|\,\cdot\,\clc}
\newcommand{\sef}{\{ \,\cdot\,|\,\cdot\,\}}
\newcommand{\mf}{\mo\,\cdot\,|\,\cdot\,\mc}
\newcommand{\cons}{{\tt cons_\Th}(\,\cdot\,,\,\cdot\,)}

\newcommand{\nat}{\mathbb{N}}
\newcommand{\dentro}{\phantom{a}}

\newcommand{\vars}{FV}

\newcommand{\Base}{\noindent\mbox{\sc Basis.}~}
\newcommand{\Step}{\noindent\mbox{\sc Step.}~}
\newtheorem{theorem}{Theorem}[section]
\newtheorem{definition}[theorem]{Definition}
\newtheorem{lemma}[theorem]{Lemma}

\newtheorem{corollary}[theorem]{Corollary}
\newtheorem{remark}[theorem]{Remark}

\newcommand{\qed}{\hfill$\Box$\par\medskip\par}
\newenvironment{proof}
      {\emph{Proof.~}~\baselineskip 8pt \small}
      {\hfill$\Box$\par\medskip\par \baselineskip 12pt
       \normalsize}
\title{A uniform approach to constraint-solving for
       lists, multisets, compact lists, and sets}
\author{{\sc Agostino Dovier}$^*$
 \and    {\sc Carla Piazza}\thanks{
         Dip. di Matematica e Informatica,
         Universit\`a di Udine.
        Via delle Scienze  206,
        33100 Udine (Italy).
        {\sf dovier$|$piazza@dimi.uniud.it}}
 \and    {\sc Gianfranco Rossi}\thanks{
        Dip. di Matematica,
        Universit\`a di Parma.
        Via M. D'Azeglio 85/A,
        43100 Parma (Italy).
       {{\sf gianfranco.rossi@unipr.it}}}
}

\begin{document}

\maketitle

\vspace*{-4ex}

\begin{abstract}
Lists, multisets, and sets are well-known data structures whose
usefulness is widely recognized in various areas of Computer
Science. These data structures have been analyzed from an
axiomatic point of view with a parametric approach
in~\cite{DPR98-fuin} where the relevant {unification algorithms}
have been developed. In this paper we extend
these results considering more general constraints including not
only equality but also membership constraints as well as their
negative counterparts.
\end{abstract}

\begin{center}
\end{center}

\section{Introduction}

Programming and specification languages usually allow the user to
represent various forms of aggregates of data objects,
characterized by the way elements are organized and accessed. In
this paper we consider four different kinds of aggregates: lists,
multisets, compact lists, and sets. The basic difference between
them lies in the order and/or repetitions of their data objects.

Importance of these forms of aggregates is widely recognized in
various areas of Computer Science. Lists are the classical example
used to introduce dynamic data structures in imperative
programming languages. They are the fundamental data structure in
functional and logic languages. Sets are the main data structure
used in specification languages (e.g., in Z~\cite{PST96}) and in
high-level declarative programming
languages~\cite{BNST91,JLP1,Ger97,HL94}; but also imperative
programming languages may take advantage from the set data
abstraction (e.g., SETL~\cite{SDDS86}). Multisets, often called
\emph{bags} in the literature, emerge as the most natural data
structure in several interesting
applications~\cite{BM93,GM96,TZO98}. A compact list is a list in
which contiguous occurrences of the same element are immaterial;
some possible application examples are suggested
in~\cite{DPR98-fuin}.

\par\smallskip\par
\begin{minipage}[b]{0.4\textwidth}
$$\begin{array}{crclc}
&&\mbox{Sets}&&\\
&\nearrow&&\nwarrow&\\
&\mbox{Multisets}&&\mbox{Compact lists}&\\
&\nwarrow &&\nearrow&\\
&&\mbox{Lists}&&\\
\end{array}$$
\vspace{5pt}
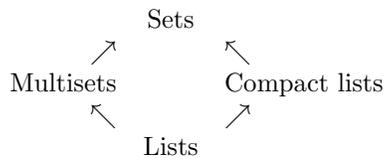
\centerline{\label{lattice}The lattice of the four
aggregates}
\end{minipage}
\phantom{aaa}
\begin{minipage}[b]{0.5\textwidth}
Lists, multisets, compact lists, and sets have been analyzed from
an axiomatic point of view and studied in the context of
(Constraint) Logic Programming (CLP)
languages~\cite{DPR98-fuin}---see figure on the left
for a lattice induced by their axiomatizations. In this context,
these aggregates are conveniently represented as terms, using
different constructors.
\end{minipage}
\par\smallskip\par

The theories studied  deal with aggregate constructor symbols as
well as with an arbitrary number of free constant and function
symbols. \cite{DPR98-fuin} focuses on \emph{equality} between
terms in each of the four theories. This amounts to solve the
unification problems in the equational theories describing the
properties of the four considered aggregates. Unification
algorithms for all of them are provided in~\cite{DPR98-fuin};
NP-unification algorithms for sets and multisets are also
presented in~\cite{ADR99,DV99}. In Section~\ref{teorie} and
\ref{sec.unif} we recall the main results of~\cite{DPR98-fuin}.

In this paper we extend the results of~\cite{DPR98-fuin} to the
case of more general \emph{constraints}. The constraints we
consider are conjunctions of literals based on both equality and
membership predicate symbols. For the case of sets, the problem
is studied in~\cite{DPPR00,DR93}. In Section~\ref{privileged} we
define the notion of constraints and we identify the privileged
models  for the axiomatic theories used to describe the
considered aggregates. We show that satisfiability of constraints
in those models is equivalent to satisfiability in any model. We
then define the notion of solved form for constraints, and we
prove that solved form constraints are satisfiable over the
proposed privileged models. In Section~\ref{constraint-rewriting}
we describe, for each kind of aggregate, the constraint rewriting
procedures used to eliminate all atomic constraints not in solved
form. We use these procedures in Section~\ref{constraint-solving}
to solve the general satisfiability problem for the considered
constraints. Some conclusions are drawn in
Section~\ref{conclusions}. Throughout the paper the word
\emph{aggregate} is used for denoting generically one of the four
considered aggregates, namely lists, multisets, compact lists,
and sets.

\section{Preliminary Notions}
\label{preliminari}

Basic knowledge of first-order logic
(e.g.,~\cite{CK73,End73}) is assumed; in this section we recall
some notions and we fix some notations that we will use throughout
the paper.

\medskip

A first-order language ${\cal L} = \langle
\Sigma, {\cal V}\rangle$ is defined by a \emph{signature} $\Sigma
= \langle {\cal F}, \Pi \rangle$ composed by a set ${\cal F}$ of
constant and function symbols, by a set $\Pi$ of predicate
symbols, and by a denumerable set $\cal V$ of variables. A
\emph{(first-order) theory} $\T$ on a language $\cal L$ is a set
of closed first-order formulas of $\cal L$ such that each closed
formula of $\cal L$ which can be deduced from $\T$ is in $\T$\/. A
\emph{(first-order) set of axioms} $\Theta$ on $\cal L$ is a set of
closed first-order formulas of $\cal L$\/. A set of axioms $\Theta$ is
said to be an \emph{axiomatization} of $\T$ if $\T$ is the
smallest theory such that $\Theta\subseteq \T$\/. Sometimes we use
the term \emph{theory} also to refer to an axiomatization of the
theory. When $\Theta = \{\varphi_1,\dots,\varphi_n\}$,
and $A_1,\dots,A_n$ are the names of the formulas
$\varphi_1,\dots,\varphi_n$, we refer to that theory simply
as: $A_1\cdots A_n$.

\medskip

Capital letters $X,Y,Z$\/, etc.\ are used to represent variables,
$f$\/, $g$\/, etc. to represent constant and function symbols, and $p$\/,
$q$\/, etc. to represent predicate symbols. We also use $\bar X$
to denote a (possibly empty) sequence of variables. $T({\cal
F},{\cal V})$ ($T({\cal F})$) denotes the set of first-order terms
(resp., ground terms) built from $\cal F$ and ${\cal V}$ (resp.,
$\cal F$). The function $\size:T({\cal F},{\cal V})\longrightarrow
\nat$ returns the number of occurrences of constant and function
symbols in a term. Given a term $t$, with $\vars(t)$ we denote
the set of all variables
which occur in the term $t$.
Given a sequence of terms $t_1,\ldots,t_n$\/,
$\vars(t_1,\ldots,t_n)$ is the set $\bigcup_{i=1}^n \vars(t_i)$.
When the context
is clear, we use $\bar t$ to denote a sequence $t_1,\ldots,t_n$ of
terms. If $\varphi$ is a first-order formula, $\vars(\varphi)$
denotes the set of free variables in $\varphi$\/. $\exists
\varphi$ ($\forall \varphi$) is used to denote the existential
(universal) closure of the formula $\varphi$\/, namely $\exists
X_1 \cdots \exists X_n\, \varphi$ ($\forall X_1 \cdots \forall X_n
\, \varphi$\/), where $\{ X_1 ,\dots ,X_n \} = \vars(\varphi)$\/.
An \emph{equational axiom} is a formula of the form $\forall
X_1\cdots\forall X_n (\ell = r)$ where $\vars(\ell=r) =
\{X_1,\dots,X_n\}$. An \emph{equational theory} is an
axiomatization whose axioms are equational axioms.

\medskip

Given a first-order theory ${\cal L} = \langle \Sigma, {\cal
V}\rangle$, a $\Sigma$-\emph{structure} is a pair $\mathcal{A} =
\langle A, I \rangle$ where $A$ is a non-empty set (the domain)
and $I$ is the interpretation function of all constant, function,
and predicate symbols of $\Sigma$ on $A$. A \emph{valuation}
$\sigma$ is a function from a subset of the set of variables
${\cal V}$ to $A$. $\sigma$ and $I$ determine uniquely a function
$\sigma^{I}$ from the set of first-order terms over $\cal L$ to
$A$ and a function from the set of formulas over $\cal L$ to the
set $\{\false,\true\}$. When the $\Sigma$-structure is fixed,
$\sigma^I$ depends only by $\sigma$. Thus, with abuse of notation,
$\sigma^I$ is simply written as $\sigma$. Given a
$\Sigma$-structure $\cal A$, a valuation $\sigma$ is said a
\emph{successful valuation} of $\varphi$ if ${\sigma}(\varphi) =
\true$. This fact is also denoted by: ${\mathcal A}\models
\sigma^{I}(\varphi)$. A formula $\varphi$ is \emph{satisfiable}
in $\mathcal{A}$ if there is a valuation
$\sigma:\vars(\varphi)\longrightarrow A$ such that $\mathcal{A}
\models \sigma(\varphi)$. In this case we say that ${\mathcal
A}\models \exists\varphi$. We say that ${\mathcal A}\models
\varphi$ if for every valuation $\sigma$ from $\vars(\varphi)
\longrightarrow A$ it holds that $\mathcal{A} \models
\sigma(\varphi)$. A formula $\varphi$ is \emph{satisfiable} in
$\mathcal{A}$ if there is a valuation
$\sigma:\vars(\varphi)\longrightarrow A$ such that $\mathcal{A}
\models \sigma(\varphi)$. In this case we say that ${\mathcal
A}\models \exists\varphi$. We remind that a formula is
satisfiable in a $\Sigma$-structure $\mathcal{A}$ if and only if
its existential closure is satisfiable in $\mathcal{A}$. Two
formulas $C_1$ and $C_2$ are \emph{equi-satisfiable} in
$\mathcal{A}$ if:  $C_1$ is satisfiable in $\mathcal{A}$ if and
only if $C_2$ is satisfiable in $\mathcal{A}$. A structure $\cal
A$ is a \emph{model} of a theory $\cal T$ if $\mathcal{A} \models
\varphi$ for all $\varphi$ in $\cal T$. We say that ${\cal T}
\models  \varphi$ if ${\cal A} \models \varphi$ for all models
$\cal A$ of $\cal T$.

\section{The Theories} \label{teorie}

For each aggregate considered, we assume that
$\Pi$ is $\{ = , \in\}$ and ${\cal F}$ contains the constant
symbol $\nil$\/ and exactly one among the binary function symbols:
$$\begin{array}{clccl}
{\lf} & \mbox{for lists,} & \phantom{aaaa} &
 \mf  & \mbox{for multisets,}\\
 \lcf & \mbox{for compact lists,}&&
 \sef  & \mbox{for sets,}
\end{array}$$
Moreover, each signature can contain an arbitrary number of other
constant and function symbols.
The four function symbols above are referred as the \emph{aggregate
constructors}. The empty list, multiset, compact list, and set are
all denoted by the constant symbol $\nil$\/. We use simple
syntactic notations for terms built using these symbols. In
particular, the list
$[\,s_1\,|\,[\,s_2\,|\,\cdots\,[\,s_n\,|\,t\,] \cdots ]]$ will be
denoted by $[s_1,\dots,s_n\,|\,t]$ or simply by $[s_1,\dots,s_n]$
when $t$ is $\nil$\/. The same conventions will be exploited also
for the other aggregates.

\subsection{Lists}\label{hybrid.lists}

The language ${\cal L}_\lst$ is defined as $\langle
\Sigma_\lst,{\cal V}\rangle$, where $\Sigma_\lst=\langle {\cal
F}_\lst,\Pi\rangle$, $\lf$ and $\nil$ are in ${\cal F}_\lst$, and
$\Pi=\{=,\in\}$\/. We recall that ${\cal
F}_\lst$ can contain other constant and function symbols.
The first-order theory \lst\ for lists is shown
in the figure below.
$${\renewcommand{\arraystretch}{\scala}
\begin{array}{|crcl|}
\hline
(K)  & \forall x\,y_1\cdots y_n & (x \not\in
     f(y_1,\dots,y_n)\,) &   f \in {\cal F}_\lst, f \mbox{ is not }
     [\,\cdot\,|\,\cdot\,]\\
(W) &  \forall y \,v \, x \,  & (x \in [\,y\,|\,v\,]
          \leftrightarrow x \in v \vee x =  y) &  \\
(F_1) & \forall x_1\cdots x_n y_1 \cdots y_n
      & \left(\begin{array}{c}
        f(x_1,\dots,x_n) =  f(y_1,\dots,y_n)\\
        \rightarrow
        x_1 =  y_1 \wedge \cdots \wedge x_n = y_n
        \end{array}\right)
        & f \in {\cal F}_\lst \\
(F_2) & \forall x_1\cdots x_m y_1 \cdots y_n
                & f(x_1,\dots,x_m) \not=  g(y_1,\dots,y_m)
         &  f,g\in{\cal F}_\lst, f \mbox{ is not } g\\
(F_3) & \forall x &
        (x \not=  t[x]) &\\
      & \multicolumn{3}{l|}{\mbox{\em where $t[x]$ denotes
      a term $t$, having $x$ as proper subterm}}\\
\hline
\end{array}}$$

The three axiom schemata $(F_{1}),(F_{2})$\/, and $(F_{3})$
(called freeness axioms, or
Clark's equality axioms---see~\cite{Cla78}) have been
originally introduced by Mal'cev in~\cite{Mal71}.
Observe that $(F_1)$ holds for
$[\,\cdot\,|\,\cdot\,]$ as a particular case.
$(F_3)$  states that there is no term
which is also a subterm of itself.
Note that $(K)$ implies that $\forall x \, ( x \notin \nil)$\/.

\subsection{Multisets}\label{sub.bags}

The language ${\cal L}_\bag$ is defined as $\langle
\Sigma_\bag,{\cal V}\rangle$, where $\Sigma_\bag=\langle {\cal
F}_\bag,\Pi\rangle$, $\mf$ and $\nil$ are in ${\cal F}_\bag$, and
$\Pi=\{=,\in\}$\/. A theory of multisets---called \bag---can be
simply obtained from the theory of lists shown above. The
constructor $[\,\cdot\,|\,\cdot\,]$ is replaced by the constructor
$\mo\,\cdot\,|\,\cdot\,\mc$ in axiom schema $(K)$ and axiom $(W)$.
The behavior of this new symbol is regulated by the following
equational axiom
$$\renewcommand{\arraystretch}{\scala}
\begin{array}{|cclr|}
\hline
  (E_p^m) & \dentro &
  \forall x y z \,\,\, \mo x,y\,|\,z\mc =  \mo y,x\,|\,z\mc &
  (\emph{permutativity})\\
\hline
\end{array}$$
which, intuitively, states that the order of elements in a
multiset is immaterial. Axiom schema $(F_1)$ does not hold for
multisets,  when $f$ is $\mo\,\cdot\,|\,\cdot\,\mc$\/. It is
replaced by axiom schemata $(F_1^{m})$:
$$\renewcommand{\arraystretch}{\scala}
\begin{array}{|clrl|}
\hline
   (F_1^m) & \dentro & \forall x_1\cdots x_n y_1 \cdots y_n
        & \left(\begin{array}{c}
           f(x_1,\dots,x_n) =  f(y_1,\dots,y_n)\\
           \rightarrow
            x_1 =  y_1 \wedge \cdots \wedge x_n = y_n
           \end{array}\right)\\
 & \multicolumn{3}{l|}{\mbox{\em for any $f \in {\cal F}_\bag$,
 $f$ distinct from $\mo\,\cdot\,|\,\cdot\,\mc$\/}}\\
\hline
\end{array}$$
In the theory $KWE_p^mF_1^mF_2F_3$\/, however, we lack in a general
criterion for establishing equality and disequality between
multisets. To obtain it, the following \emph{multiset
extensio\-na\-li\-ty\/} property is introduced: \emph{Two
multisets are equal if and only if they have the same number of
occurrences of each element, regardless of their order.} The axiom
proposed in~\cite{DPR98-fuin} to force this property is the
following:
$$
\renewcommand{\arraystretch}{\scala}
\begin{array}{|cll|}
\hline
(E^m_k) &\dentro &
   \forall  y_1 y_2 v_1 v_2\,
   \left(\begin{array}{l}
   \mo y_1\,|\,v_1 \mc =  \mo y_2\,|\,v_2 \mc \:\:
   \leftrightarrow \\
   \phantom{aaaaaaaa}
   (y_1 =  y_2 \wedge v_1 =  v_2) \vee\\
   \phantom{aaaaaaaa}
   \exists z\,(v_1 =  \mo y_2\,|\,z \mc \wedge
                          v_2 =  \mo y_1\,|\,z \mc )
   \end{array}\right)\\
\hline
\end{array}$$
$(E^m_k)$ implies $(E^m_p)$\/.
Axiom schema $(F^m_3)$ is also introduced:
$$
\renewcommand{\arraystretch}{\scala}
\begin{array}{|cll|}
\hline
(F_3^m) &
   \forall  x_1\cdots x_m y_1 \cdots y_n x\, &
   \left( \begin{array}{l}
      \mo x_1,\dots,x_m \,|\, x \mc =
      \mo y_1,\dots,y_n\,|\,x \mc \\
      \rightarrow \mo x_1,\dots,x_m  \mc =
      \mo y_1,\dots,y_n \mc
      \end{array} \right)  \\
\hline
\end{array}$$

Axiom schema $(F_3^m)$ reinforces the acyclicity condition
imposed by standard axiom schema $(F_3)$. As a matter of fact, $X
\neq \mo a , b , b \,|\, X\mc$ follows from $(F_3)$. Axiom schema
$(F_3^m)$ states that, since $\mo a , a , b \mc \neq \mo a , b ,b
\mc$, then $\mo a , a , b\,|\,X \mc \neq \mo a , b ,b \,|\,X
\mc$. This property is not a consequence of the the remaining
part of the theory.

\subsection{Compact Lists}\label{compact}

The language ${\cal L}_\clist$ is defined as ${\cal
L}_\clist=\langle \Sigma_\clist,{\cal V}\rangle$, where
$\Sigma_\clist=\langle {\cal F}_\clist,\Pi\rangle$, $\lcf$ and
$\nil$ are in ${\cal F}_\clist$, and $\Pi=\{=,\in\}$\/. Similarly
to multisets, the theory of \emph{compact lists}---called
\clist---is obtained from the theory of lists with only a few
changes. The list constructor symbol is replaced by the binary
compact list constructor $\clo\,\cdot\,|\,\cdot\,\clc$ in $(K)$
and $(W)$. The behavior of this symbol is regulated by the
equational axiom
$$\renewcommand{\arraystretch}{\scala}
\begin{array}{|cclr|}
\hline
  (E_a^c) & \dentro &
  \forall x y   \,\,\, \clo x,x\,|\,y\clc =  \clo x\,|\,y\clc &
  (\emph{absorption}) \\
\hline
  \end{array}$$
which, intuitively, states that contiguous duplicates in a
compact list are immaterial.
As for multisets, we introduce a general criterion for establishing
both equality and disequality between compact lists.
This is obtained by introducing the following axiom:
$$\renewcommand{\arraystretch}{\scala}
\begin{array}{|clll|}
\hline
(E^c_k) &\dentro &
   \forall  y_1 y_2 v_1 v_2 &
   \left(\begin{array}{l}
   \clo y_1\,|\,v_1 \clc =  \clo y_2\,|\,v_2 \clc \:\:
   \leftrightarrow \\
   \phantom{aaaaaaaa}
   (y_1 =  y_2 \wedge v_1 =  v_2) \vee\\
   \phantom{aaaaaaaa}
   (y_1 =  y_2 \wedge v_1 =  \clo y_2\,|\,v_2 \clc) \vee\\
   \phantom{aaaaaaaa}
   (y_1 =  y_2 \wedge \clo y_1\,|\,v_1 \clc =  v_2) \\
   \end{array}\right) \\
\hline
\end{array}$$
$(E^c_a)$ is implied by $(E^c_k)$.
Axiom schema $(F_1)$ is replaced by axiom schema $(F_1^c)$:
$$\renewcommand{\arraystretch}{\scala}
\begin{array}{|clrl|}
\hline
   (F_1^c) & \dentro & \forall x_1\cdots x_n y_1 \cdots y_n
        & \left(\begin{array}{c}
           f(x_1,\dots,x_n) =  f(y_1,\dots,y_n)\\
           \rightarrow
            x_1 =  y_1 \wedge \cdots \wedge x_n = y_n
           \end{array}\right)\\
 & \multicolumn{3}{l|}{\mbox{\em for any $f \in {\cal F}_\clist$,
 $f$ distinct from $\clo\,\cdot\,|\,\cdot\,\clc$\/}}\\
\hline
\end{array}$$
The freeness axiom $(F_3)$ needs to be
suitably modified. The introduction of $(F_3)$ is motivated by
the requirement of finding solutions to equality constraints
over $\Sigma$-structures with the domain built based on
Herbrand Universe, where each term is
modeled by a finite tree. As opposed to lists and multisets, an
equation such as
$X =  \clo \nil\,|\, X\clc$ admits a solution in these structures.
Precisely, a solution that binds $X$ to the term
$\clo \nil\,|\,t \clc$, where $t$ is any term.
Therefore, as explained in~\cite{DPR98-fuin},
axiom schema $(F_3)$ should be weakened and, thus, replaced by:
$$
\renewcommand{\arraystretch}{\scala}
\begin{array}{|ccll|}
\hline
 (F_3^c) & \dentro & \forall x & (x \neq  t[x])\\
 & \multicolumn{3}{l|}{\mbox{\em unless: $t$ is of the form
      $\clo t_1,\dots,t_n\,|\,x \clc$\/, with $n > 0$,}}\\
 & \multicolumn{3}{l|}{\mbox{\em $x \notin \vars(t_1,\dots,t_n)$\/, and $t_1=  \cdots = t_n$}}\\
\hline
\end{array}$$

\subsection{Sets}\label{Sets}

The language ${\cal L}_\set$ is defined as
${\cal L}_\set=\langle \Sigma_\set,{\cal V}\rangle$, where
$\Sigma_\set=\langle {\cal
F}_\set,\Pi\rangle$,
$\sef$
and $\nil$
are in ${\cal F}_\set$, and
$\Pi=\{=,\in\}$\/.
The last theory we consider is the simple theory of sets \set.
Sets have both the \emph{permutativity\/} and the
\emph{absorption properties\/}
which, in the case of $\sef$\/,
can be rewritten as follows:
$$\renewcommand{\arraystretch}{\scala}
\begin{array}{|ccrcl|}
\hline
  (E_p^s) & \dentro &
  \forall x y z \,\,\, \{ x,y\,|\,z\} & =  & \{ y,x\,|\,z\}
\\
  (E_a^s) & \dentro &
  \forall x y   \,\,\, \{ x,x\,|\,y\} & =  & \{ x\,|\,y\}\\
\hline
  \end{array}$$
A criterion for testing equality (and disequality)
between sets is obtained by merging
the multiset equality axiom $(E^m_k)$ and the compact
list equality axiom $(E^c_k)$\/:
$$\renewcommand{\arraystretch}{\scala}
\begin{array}{|clll|}
\hline
(E^s_k) &\dentro &
   \forall  y_1 y_2 v_1 v_2 &
   { \left(\begin{array}{l}
   \{ y_1\,|\,v_1 \} =  \{ y_2\,|\,v_2 \} \:\:
   \leftrightarrow \\
   \phantom{aaaaaaaa}
   (y_1 =  y_2 \wedge v_1 =  v_2) \vee\\
   \phantom{aaaaaaaa}
   (y_1 =  y_2 \wedge v_1 =  \{ y_2\,|\,v_2 \}) \vee\\
   \phantom{aaaaaaaa}
   (y_1 =  y_2 \wedge \{ y_1\,|\,v_1 \} =  v_2) \vee\\
   \phantom{aaaaa}\exists k \:( v_1 =  \{y_2\,|\,k\} \wedge
                 v_2 =  \{y_1 \,|\, k \} )
   \end{array}\right) }\\
\hline
\end{array}
$$
According to $(E^s_k)$ duplicates and ordering of elements in
sets are immaterial. Thus, $(E^s_k)$ implies the equational axioms
$(E^s_p)$ and $(E^s_a)$\/. In~\cite{DPR98-fuin} it is also proved
that they are equivalent when domains are made by terms. The
theory \set\ also contains axioms $(K)$\/, $(W)$ with
$[\,\cdot\,|\,\cdot\,]$ replaced by $\sef$\/, and axiom schemata
$(F_2)$ Axiom schema $(F_1)$ is replaced by:
$$\renewcommand{\arraystretch}{\scala}
\begin{array}{|clrl|}
\hline
   (F_1^s) & \dentro & \forall x_1\cdots x_n y_1 \cdots y_n
        & \left(\begin{array}{c}
           f(x_1,\dots,x_n) =  f(y_1,\dots,y_n)\\
           \rightarrow
            x_1 =  y_1 \wedge \cdots \wedge x_n = y_n
           \end{array}\right)\\
 & \multicolumn{3}{l|}{\mbox{\em for any $f \in {\cal F}_\set$,
 $f$ distinct from $\{\,\cdot\,|\,\cdot\,\}$\/}}\\
\hline
\end{array}$$
The modification of axiom schema $(F_3)$ for sets, instead,
simplifies the one used for compact lists:
$$
\renewcommand{\arraystretch}{\scala}
\begin{array}{|ccll|}
\hline
(F_3^s) & \dentro & \forall x & (x \not=  t[x])\\
& \multicolumn{3}{l|}{\mbox{\em unless: $t$ is of the form
     $\{ t_1,\dots,t_n\,|\,x \}$ and
     $x \in \vars(t_1,\dots,t_n)$}}\\
\hline
\end{array}$$

\subsection{Equational theories}\label{Sunto}

As we have seen in this section, each aggregate constructor is
precisely characterized by zero, one or 2 equational axioms. We
define the four corresponding \emph{equational theories} as
follows:
\begin{center}
\begin{tabular}{|rl|}
\hline
$E_\lst$ & the empty theory for \lst,\\
$E_\bag$ &  consisting of the \emph{Permutativity}
axiom $(E^m_p)$ for \bag, \\
$E_\clist$ &  consisting of the \emph{Absorption}
axiom $(E^c_a)$ for \clist,\\
$E_\set$ &  consisting of both the
\emph{Permutativity} $(E^s_p)$ and
\emph{Absorption} $(E^s_a)$ axioms for \set.\\
\hline
\end{tabular}
\end{center}
Relationships between these equational theories,
$\Sigma$-structures, and the proposed first-order theories for
aggregates are explained in the next section.
Figure~\ref{sommario} summarizes the axiomatizations of the four
theories.

\begin{figure}[t]
\begin{center}
{\small
\begin{tabular}{c|c|c|c|c|c|c||c|c|c}
Name & empty & with & \multicolumn{2}{|c|}{Equality} &
Herbrand & Acycl. & Perm. & Abs. & Equational Name \\
\hline
\lst & $(K)$ & $(W)$ & \multicolumn{2}{|c|}{$(F_1)$} &
$(F_2)$ & $(F_3)$ & & & $E_\lst$ \\
\hline
\bag & $(K)$ & $(W)$ & $(E^m_k)$ & $(F^m_1)$ &
$(F_2)$ & $(F_3)$ & $(E^m_p)$ & & $E_\bag$\\
\hline
\clist & $(K)$ & $(W)$ & $(E^c_k)$ & $(F^c_1)$ &
$(F_2)$ & $(F^c_3)$ & & $(E^c_a)$ & $E_\clist$\\
\hline
\set & $(K)$ & $(W)$ & $(E^s_k)$ & $(F^s_1)$ &
$(F_2)$ & $(F^s_3)$ & $(E^s_p)$ & $(E^s_a)$ & $E_\set$\\
\hline
\end{tabular}}
\end{center}
\caption{\label{sommario}Axioms for the four theories}
\end{figure}

\section{Constraints, Privileged Models, and Solved Form}
\label{privileged}

In this section we introduce the privileged models for the
four theories introduced in the previous section. These models
are used to testing satisfiability of the particular kind of
formulas we are concerned with, namely, \emph{constraints}. We
then show that the models and the theories defined in the
previous section \emph{correspond} on the class of constraints
considered. Moreover, we give a general notion of \emph{solved
form} for constraints, and we prove that a solved form constraint
is satisfiable in the corresponding privileged model.

\begin{definition}[Constraints]  \label{constraintdef}
Let $\Th$ be either $\lst$ or $\bag$ or $\clist$ or $\set$\/. A
\emph{$\Th$-constraint} $C_\Th$ is a conjunction of atomic ${\cal
L}_\Th$-formulas or negation of atomic ${\cal L}_\Th$-formulas of
the form $s \mathbin{\pi} t$,
where $\mathbin{\pi}\in \Pi$\/,
and $s,t \in T({\cal F}_{\Th},{\cal V})$.
\end{definition}

Throughout the paper we will use the following terminology to
refer to particular kinds of constraints: \emph{equality} (resp.,
\emph{disequality}) \emph{constraints} are conjunctions of atomic
formulas of the form $s = t$ (resp., $s \neq t$).
\emph{Membership} (resp., \emph{not-membership}) constraints are
conjunctions of membership atoms (resp., membership negative
literals), i.e. formulas of the kind $s \in t$  (resp., $s\not\in
t$).

\subsection{Privileged Models}\label{privistru}

As discussed in Section~\ref{Sunto}, each aggregate constructor
is precisely characterized by an equational theory, that we have
named $E_\lst$, $E_\bag$, $E_\clist$, and $E_\set$. Using the
appropriate equational theory we can define a privileged model
for the first-order theory $\lst$, $\bag$, $\clist$, and $\set$
for each aggregate. Each model is obtained as a partition of the
Herbrand Universe.

\begin{definition}\label{defstruttura}
Let $\Th$ be $\lst$ (resp., $\bag$, $\clist$, or $\set$\/). A
\emph{privileged $\Sigma$-structure for $\Th$} is defined
as follows.
\begin{enumerate}
\item The \emph{domain} of the $\Sigma$-structure is the quotient
$T({\cal F}_\Th)/\equiv_\Th$ of the Herbrand Universe
$T({\cal F}_\Th)$
over the smallest congruence relation $\equiv_\Th$ induced
by the equational theory $ E_\Th$ on  $T({\cal F}_\Th)$\/.
\item The interpretation of a term $t$ is its equivalence
   class w.r.t.\ $\equiv_\Th$, denoted by $\rappr{t}$\/.

\item $=$ is interpreted as the identity on the
domain $T({\cal F}_\Th)/\equiv_\Th$\/.

\item The interpretation of membership is:
   $ \rappr{t} \in \rappr{s} $ is \true\ if and only if there
   is a term in $\rappr{s}$ of the form
      $[ t_1,\dots,t_n ,  t\,|\,r ]$
(resp., $\mo   t_1,\dots,t_n ,  t\,|\,r \mc$\/,
$\clo  t_1,\dots,t_n ,  t\,|\,r  \clc$\/,
or $ \{ t_1,\dots,t_n ,  t\,|\,r \}$\/)
for some terms $t_1,\dots,t_n,r$\/.
\end{enumerate}
\end{definition}

It is easy to prove that the above defined $\Sigma$-structures
are in fact models of the corresponding theories.
In Lemma~\ref{versofacile} we prove this property for multisets.
>From now on, we will call the privileged $\Sigma$-structures above
defined \emph{privileged models} for $\lst$, $\bag$, $\clist$, and
$\set$. We refer to them as
$\cal LIST$, $\cal MSET$, $\cal CLIST$, and $\cal SET$\/,
respectively.

\begin{remark}\label{member set}
When $\rappr{s}$ is the class of a multiset (resp.,  a set),
since the permutativity property holds, the requirement for $
\rappr{t} \in \rappr{s} $ to be \true\ can be simplified to: $\mo
t\,|\,r \mc$ (resp., $ \{ t\,|\,r \}$) is in $ \rappr{s}$.
\end{remark}

The following notion from~\cite{JM94}
is crucial for characterizing the above
privileged models.

\begin{definition}
Given a first-order language ${\cal L} = \langle \Sigma,
{\cal V} \rangle$,
a set of first-order formulas $\cal C$ on $\cal L$,
a theory $\T$ on $\cal L$, and a
$\Sigma$-structure ${\cal A}$\/,
$\T$ and $\cal A$ \emph{correspond} on the
set $\cal C$ if, for each $\varphi \in \cal C$\/, we have that $\T
\models  \exists \varphi$ if and only if ${\cal A} \models
\exists \varphi$\/.
\end{definition}

This property means that if $\varphi$ is an element of $\cal C$
and $\varphi$ is satisfiable in $\cal A$, then it is satisfiable
in all the models of $\T$\/. We prove the correspondence property
for our theories and the privileged models, when the class $\cal
C$ is the class of constraints defined in Definition
\ref{constraintdef}. We show below the proof of this result in the
case of the model $\cal MSET$ and the theory \bag. The other
cases are similar. In the proof we use some basic results which
can be found in the Appendix~\ref{modellistiche}
(Lemmas~\ref{logica}--\ref{versodifficile}).

\begin{theorem}\label{corrispondenza}
The model $\cal MSET$ (resp., $\cal LIST$, $\cal CLIST$, $\cal
SET$\/) and the theory  \bag\ (resp., \lst, \clist, and \set)
correspond on the class of \bag- (resp., \lst-, \clist-, and
\set-)constraints.
\end{theorem}
\begin{proof}
{F}rom Lemma~\ref{versofacile} it follows that
${\cal MSET}$ is a model of \bag, namely that
if $C$ is a
first-order formula and $\bag\models C$\/, then ${\cal
MSET}\models C$\/.

On the other hand, if $\exists C$ is a formula with only
existential quantifiers, then ${\cal MSET}\models \exists C$
if and only if there exists $\sigma$ such that ${\cal MSET}\models
\sigma(C)$\/.
Assume that ${\cal M}\models \sigma(C)$. From Lemmas~\ref{logica}
and~\ref{versodifficile}, we have that ${\cal M}\models
\exists C$ for all models ${\cal M}$ of \bag. This implies that
$\bag \models \exists C$\/.
\end{proof}

\subsection{Solved Form}

Solved form constraints play a fundamental r\^ole in establishing
satisfiability of constraints in the corresponding privileged
model. The solved form is obtained by defining first a weaker
form, called the pre-solved form, and then by adding to this form
two further conditions.

\begin{definition}
\label{pre-solvedform}
A constraint $C = c_1 \wedge \cdots \wedge c_n$ is in
\emph{pre-solved form} if for $i \in \{1,\dots,n\}$, $c_i$ is in
\emph{pre-solved form in $C$}, i.e.\ in one of the following forms:
\begin{itemize}
\item $X = t$ and $X$ does not occur elsewhere in $C$
\item $t \in X$ and $X$ does not occur in $t$
\item $X \neq t$ and $X$ does not occur in $t$
\item $t \notin X$ and $X$ does not occur in $t$\/.
\end{itemize}
\end{definition}

A constraint in pre-solved form is not guaranteed to be
satisfiable in the corresponding privileged model. For example, the
constraint $X \in Y \wedge Y \in X$ is in pre-solved form but it
is unsatisfiable in each of the privileged models $\cal LIST, MSET, CLIST$\/, and
$\cal SET$. The first condition we introduce below takes care of
this situation.

\begin{definition}[Acyclicity Condition] Let $C$ be a pre-solved
form constraint and $C^{\in}$ be the part of $C$ containing only
membership constraints. Let ${\cal G}^{\in}_{C}$ be the directed
graph obtained as follows:
\begin{description}
\item[\emph{Nodes}.] Associate a distinct node to
  each variable $X$ in $C^{\in}$\/.
\item[\emph{Edges}.] If $t \in X$ is in $C^{\in}$\/,
     $\nu_1,\dots,\nu_n$ are the nodes
     associated with the variables in $t$\/, and
$\mu$  is the node associated with the variable
$X$\/, then add the  edges
$\langle \nu_1,\mu\rangle,\dots,\langle\nu_n,\mu\rangle$\/.
\end{description}
We say that a pre-solved form constraint $C$ is \emph{acyclic} if
${\cal G}_{C}^{\in}$ is acyclic.
\end{definition}

The acyclicity condition is not sufficient for satisfiability.
Consider the constraint $\{ A,B \} \in X \wedge \{ B,A \} \notin
X$. It is in pre-solved form and acyclic but unsatisfiable in all
the considered privileged models. Conversely, the constraint $\{
A \} \in X \wedge \{ a \} \notin X$ is satisfiable in $\cal SET$
(e.g., $A = b, X = \{\{b\}\}$). We observe that whenever there
are two constraints $t \in X$ and $t' \not\in X$ in $C$ such that
$t$ and $t'$ are equivalent terms in the equational theory
$E_\Th$, the constraint $C$ is unsatisfiable.

This analysis, however, does not cover all the possible cases in
which an acyclic constraint in pre-solved form is unsatisfiable,
as it ensues from the following example:
$$a\in X\wedge X\in Y\wedge \{a\,|\,X\}\not\in Y.$$
Observe that there are no pairs of terms $t,t'$ of the form
singled out above. Nevertheless, since the satisfiability of $a
\in X$ is equivalent in \set\ to that of $X = \{a\,|\,N\}$ ($N$ is
a new variable), we have that the constraint is equi-satisfiable
to:
$$X = \{a\,|\,N\} \wedge  \{a\,|\,N\} \in Y\wedge
\{a,a\,|\,N\}\not\in Y.$$
Now, $\{a\,|\,N\}$ and
$\{a,a\,|\,N\}$ are equivalent terms in $E_\set$, and thus the
constraint is unsatisfiable.

\medskip

To formally define the second condition for solved form
constraints, taking into account all the possible cases informally
described above, we introduce the following definitions.

\begin{definition}
Let $\theta\equiv[X_1/t_1,\dots,X_n/t_n]$ be a substitution and
$m\in\nat$. We recursively define the substitution
$\theta^m$ as:
$$\left\{\begin{array}{rcll}\theta^1 & = & \theta\\
\theta^{m+1} & = &[X_1/\theta^m(t_1),\dots, X_n/\theta^m(t_n)] & m>0
  \end{array}\right.$$
If there exists $m>0$ such that $\theta^{m+1}\equiv \theta^{m}$
we say that $\theta$ is \emph{stabilizing}. Given a stabilizing
substitution $\theta$, the \emph{closure} $\theta^*$ of $\theta$
is the substitution $\theta^m$ such that $\forall k>m$ we have that
$\theta^{k}\equiv\theta^m$\/.
\end{definition}

\begin{definition}
\label{subC}
Let $C$ be a constraint in pre-solved form over the language
${\cal L}_{\lst}$ (${\cal L}_\bag, {\cal L}_\clist, {\cal
L}_\set$\/) and let
$t_1^1\in X_1,\dots,t_1^{k_1}\in X_1,\dots,
  t_q^1\in X_q,\dots,t_q^{k_q}\in X_q$
be all membership atoms of $C$\/. We define the \emph{member
substitution} $\sigma_C$ as follows:
$$\sigma_C\equiv[X_1/[F_1,t_1^1,\dots, t_1^{k_1}\,|\,M_1],\dots,
X_q/[F_q,t_q^{1},\dots,t_q^{k_q}\,|\,M_q] ]$$ (respectively,
$\sigma_C \equiv [ X_1/\mo F_1, t_1^1,\dots,
t_1^{k_1}\,|\,M_1\mc,\dots
],$
$\sigma_C\equiv [ X_1/ \clo
F_1,t_1^1,\dots,t_1^{k_1}\,|\,M_1\clc,\dots
],$\\
$\sigma_C\equiv [ X_1/\{ F_1, t_1^1,\dots,
t_1^{k_1}\,|\,M_1\},\dots
])$
where $F_i$ and $M_i$ are new variables not occurring in $C$\/.
\end{definition}

The member substitution $\sigma_C$ forces all the terms
$t_i^j$'s to be member of the aggregate represented by $X_i$.
The variable $F_i$ in $X_i$ is
necessary in the case of compact lists. As a matter of fact, in
every valuation $\sigma$ satisfying the constraint:
$$Y\in X_1\wedge \clo Y\,|\,X_1\clc\in X_2\wedge X_1\not\in X_2$$
it must be $\sigma(X_1)\neq\sigma(\clo Y\,|\,X_1\clc)$. Thus, in
$\sigma_C$ we give the possibility to the first element of
$\sigma(X_1)$ to be different from $\sigma(Y)$. We show in the
Appendix~\ref{modellistiche} that if $C$ is a constraint in
pre-solved form and acyclic, then $\sigma_C$ is stabilizing
(Lemma~\ref{convergente}).

We are now ready to state the second condition for the solved
form.

\begin{definition}[Membership Consistency Condition]
Let $E_\Th$ be one of the four equational theories for aggregates.
A constraint $C$ in pre-solved
form and acyclic is \emph{membership consistent} if for each pair
of literals of the form $t\not\in X, t'\in X$ in $C$ we have that:
$$E_\Th\not\models \forall (\sigma_C^*(t)=\sigma_C^*(t')).$$
\end{definition}

The definition of solved form, therefore, can be given simply as
follows:

\begin{definition}[Solved Form]\label{solvedform}
A constraint $C$ in pre-solved form is said to be in \emph{solved
form} if it satisfies the membership consistent condition.
\end{definition}

Observe that the membership consistency condition implies the
acyclicity condition. It is a semantic requirement of equivalence
of two terms under a given equational theory. However, this test
can be automatized in the following way. As well-known from
unification theory (see, e.g.,~\cite{BN98,Sie90}), given an
equational theory $E$, knowing whether two terms are equivalent
modulo $\equiv_E$ is the same as verifying whether the two terms
$t$ and $t'$ are $E$-unifiable with empty m.g.u. ($\varepsilon$).
Thus, the test is connected with the availability of a
unification algorithm for the theory $ E_\Th$\/.
In~\cite{DPR98-fuin} it is proved that the four equational
theories we are dealing with are finitary (i.e., they admit a
finite set of mgu's that covers all possible unifiers) and,
moreover, the unification algorithms for the four theories are
presented. This give us a decision procedure for the above test.

As an example, let $C$ be the pre-solved form and acyclic
\set-constraint:
$a\in Y\wedge Y\in X\wedge X\in Z\wedge \{\{a\,|\,Y\}\,|\,X\}
\not\in Z$.
It holds that:
$$\begin{array}{rcl}
\sigma_C & = & [Y/\{F_Y,a\,|\,M_Y\}, X/\{F_X,Y\,|\,M_X\},
Z/\{F_Z,X\,|\,M_Z\}]\,,\\
\sigma_C^* & = &[Y/\{F_Y,a\,|\,M_Y\},
X/\{F_X,\{F_Y,a\,|\,M_Y\}\,|\,M_X\},\\
&& \,\,
Z/\{F_Z,\{F_X,\{F_Y,a\,|\,M_Y\}\,|\,M_X\}\,|\,M_Z\}] \\
\sigma_C^*(X) &=& \{F_X,\{F_Y,a\,|\,M_Y\}\,|\, M_X\}  \\
\sigma_C^*(\{\{a\,|\,Y\}\,|\,X\}) &=&
\{\{a,F_Y,a\,|\,M_Y\},F_X,\{F_Y,a\,|\,M_Y\}\,|\,M_X\}
\end{array}$$
The
constraint is not in solved form since $E_\set\models
\forall(\sigma_C^*(X)= \sigma_C^*(\{\{a\,|\,Y\}\,|\,X\}))$\/.

\medskip

We prove now that solved form constraints are satisfiable in the
corresponding privileged models.
We prove the property for  \set-constraints.
The proof is similar for the other cases.

\begin{theorem}[Satisfiability of the Solved Form]\label{all solved}
Let $C$ be a constraint in solved form over the language ${\cal
L}_\set$ (resp., ${\cal L}_\lst$, ${\cal L}_\bag$, ${\cal
L}_\clist$). Then ${\cal SET}\models \exists C$\/ (resp., $\cal
LIST$, $\cal MSET$, and $\cal CLIST$).
\end{theorem}
\begin{proof}
We split $C$ into the four parts: $C^=$, $C^{\in}$, $C^{\notin}$,
and $C^{\neq}$, containing $=,\in, \notin$\/, and $ \neq$
literals, respectively. For all pairs of literals $p \in V,r
\notin V$ in $C$ let $NEQ_{pr}$ be an auxiliary variable, that
will be used as a `constraint store', initialized to the empty
set $\e$\/. We will use the two auxiliary functions {\it rank}
and {\it find}. The {\it rank}\/ of a well-founded set is
basically the maximum nesting of braces needed to write it.
Precisely:
$$ {\it rank}(s) =  \left\{
\begin{array}{ll}
        0 & \mbox{if $s$ is not of the form  $\{u\,|\,v\}$}\\
        \max\{1 + {\it rank}(u), {\it rank}(v)\} &
        \mbox{if $s$ is $\{u\,|\,v\}$\/}
\end{array}
\right.$$
${\it find}(X,t)$ is a function that produces for each
pair $(X,t)$ a set of integer numbers indicating the `depth' of
the occurrences of the variable $X$ in $t$\/. It can be defined
as:
$$ {\it find}(X,t) =  \left\{
\begin{array}{ll}
   \emptyset & \mbox{if $t$ is a constant term}\\
   \{0\} & \mbox{if $t$ is a variable $X$}\\
   {\{1 + n : n \in {\it find}(X,y)\}}  &
         \mbox{if $t$ is $\{y\,|\,f(t_1,\dots,t_m)\}$,
         $f$ is not $\sef$}  \\
   {\{1 + n : n \in {\it find}(X,t_1) \cup \cdots\cup
         {\it find}(X,t_m) \}}  &
         \mbox{if $t$ is $f(t_1,\dots,t_m)$, $f$ is not $\sef$} \\
   {\{1 + n : n \in {\it find}(X,y)\}\:
         \cup{\it find}(X,s)} &
         \mbox{if $t$ is $\{y\,|\,s\}$\/, $s \neq \nil$}\\
\end{array}
\right.$$
 We build a successful valuation $\gamma$ of $C$, in
various steps.

\begin{description}
\item[$C^{=}$] is of the form $ X_1 = t_1 \wedge \cdots
\wedge X_m = t_m$.
We define the mapping:
$\theta_1 = [X_1 / t_1,\dots,X_m /  t_m].$

\item[$C^{\in}$] is of the form $p_{1}^{1}\in V_{1}\wedge\cdots\wedge
p_{1}^{v1}\in V_{1}\wedge\cdots\wedge p^{vq}_{q}\in V_{q}$\/.
Consider the member substitution
$$\sigma_C = [V_{1} /
\{F_{1},p_{1}^{1},\dots,p_{1}^{v1}\,|\, M_{1}\},\dots, V_{q} /
\{F_q,p_{q}^{1},\dots,p_{q}^{vq}\,|\, M_{q}\}].$$ Since,  by
hypothesis, $C$ is acyclic, then $\sigma_C^*$ can be computed
(see Lemma~\ref{convergente}).

For each pair of literals $p \in V$, $r \notin V$ of $C$ consider the
equality constraints in solved form $D_{1},\dots,D_{k}$ that are the
solutions to the unification problem
$\sigma_C^*(p)=\sigma_C^*(r)$
(since $C$ is in solved form they are all different from the
empty substitution).
By the results concerning unification (cf.~\cite{DPR98-fuin})
we have that
$$ \sigma_C^*(p)=\sigma_C^*(r) \leftrightarrow
\bigvee_{j=1}^{k}(\exists \bar N (D_{j})),$$
where $\bar N$ are new variables, and each $D_{j}$ is a
conjunction of equations which contains at least one atom of the
form $A=\{a_{1},\dots,a_{h}\,|\,B\}$ with $A\in
FV(\sigma_C^*(p))\cup FV(\sigma_C^*(r))$ and $FV(a_{i})\subseteq
FV(\sigma_C^*(p))\cup FV(\sigma_C^*(r))$\/, or one atom of the
form $A=B$ with $A,B\in FV(\sigma_C^*(p))\cup FV(\sigma_C^*(r))$\/.

Since we want to satisfy $\sigma_C^*(r)\notin \sigma_C^*(V)$ we
are interested in satisfying
$\sigma_C^*(r)\neq \sigma_C^*(p)$, which is in turn equivalent to:
$$\bigwedge_{j=1}^{k}(\forall N\neg D_{j}).$$
For doing that, for each $D_{j}$ we choose an atom of the form
$A=\{a_{1},\dots,a_{h}\,|\,B\}$ or $A=B$ and we store it in the
variable $NEQ_{pr}$\/. Points (5) and (6) below will take care of
this constraint store.

\item[$C^{\notin}$] is of the form  $r_1 \notin Y_1 \wedge \cdots
\wedge r_n \notin  Y_n$ (\/$Y_i$ does not occur in $r_i$\/)
and
$C^{\neq}$ is of the form $Z_1 \neq  s_1 \wedge\cdots\wedge Z_o
\neq s_o$ (\/$Z_i$ does not occur in $s_i$\/).
Let $W_1,\dots,W_h$ be the variables occurring in $C$ other than
$X_1,\dots,X_m,V_{1},\dots,
V_{q},Y_1,\dots,Y_n,Z_1,\dots,Z_o$\/.

Let $\bar{s} = \max\{ {\it rank}(t): \mbox{ $t$ occurs in
$\sigma_C^*(\theta_1(C))$}\}+1+h$\/.

Let $R_1,\dots,R_j$ be the variables occurring in
$\sigma_C^*(\theta_1(C^{\notin}\wedge C^{\neq}))$ (actually, the
variables $\bar F, \bar M$\/, and some of the $\bar Y$ and $\bar
Z$\/) and $n_1,\dots,n_j$ be auxiliary variables ranging over
$\nat$\/.

We build an integer disequation system $S$ in the following way:

\begin{enumerate}

\item $S = \{n_i > \bar{s}: \forall i \in \{1,\dots,j\}\} \cup
           \{n_{i_1} \neq  n_{i_2}: \forall i_1,i_2 \in\{1,\ldots,j\},
           i_1 \neq  i_2 \}$.

\item For each literal $R_{i_1} \neq  t$
in $\sigma_C^*(C^{\neq })$
     $$S = S \cup \{ n_{i_1} \neq  n_{i_2} + c: \forall i_2 \neq
     i_1, \forall c \in {\it find}(R_{i_2},t)\}$$

\item For each literal
$\{R_{i_1},p_{j}^{1},\dots,p_{j}^{vj}\,|\,R_{h}\}\neq t$ in
$\sigma_C^*(C^{\neq })$
     $$S = S \cup \{ n_{i_1} \neq  n_{i_2} + c-1: \forall i_2 \neq
i_1, \forall c \in {\it find}(R_{i_2},t)\}$$

\item For each  literal
$t \notin  R_{i_1}$ in $\sigma_C^*(C^{\notin})$ $$    S = S \cup
\{n_{i_1} \neq  n_{i_2} + c +1:  \forall i_2 \neq  i_1, \forall c
\in {\it find}(R_{i_2},t)\}$$

\item For each literal
$t\notin \{R_h,p_{j}^{1},\dots,p_{j}^{vj}\,|\,R_{i_{1}}\}$\/,
for each
$k\leq vj$, for all $R_{i_{2}}=\{a_{1},\dots, a_{h}\,|\,B\}$
in
$NEQ_{p_{j}^{k}t}$
      $$S = S\cup \{n_{i_2} \neq  n_{i_3} + c + 1: \forall
i_3 \neq
i_2, \forall c \in {\it find}(R_{i_3},a_{1})\}$$

\item For each literal
$t\notin \{R_h,p_{j}^{1},\dots,p_{j}^{vj}\,|\,R_{i_{1}}\}$\/,
for each
$k\leq vj$, for all $R_{i_{2}}=R_{i_{3}}$ in
$NEQ_{p_{j}^{k}t}$
      $$S = S\cup \{n_{i_2} \neq  n_{i_3}\}$$

\item For each literal
$t\notin \{R_{i_1},p_{j}^{1},\dots,p_{j}^{vj}\,|\,R_{h}\}$\/,
      $$S = S\cup \{n_{i_1} \neq  n_{i_2}+c: \forall i_2\neq i_1,
      \forall c\in {\it find}(R_{i_2},t)\}$$

\item For each literal
$t\notin \{R_h,p_{j}^{1},\dots,p_{j}^{vj}\,|\,R_{i_{1}}\}$\/,
      $$S = S\cup \{n_{i_1} \neq  n_{i_2} + c + 1: \forall
i_2 \neq
i_1, \forall c \in {\it find}(R_{i_2},t)\}$$

\end{enumerate}
\end{description}

An integer disequation is \emph{safe} if, after expression
evaluation, it is not of the form $u \neq u$\/. A safe disequation
has always an infinite number of solutions. A finite set of safe
disequations has always an infinite number of solutions. We show
that all disequations of $S$ are safe. The disequations generated
at point $(1)$ are safe by definition; those introduced in points
$(2),(4),(5),(6),(7),$ and $(8)$ are safe since $c$ is always a
positive number. We prove that the disequations generated at point
$(3)$ are safe. If in $C$ there was a situation of the form
$p_1\in Y\wedge\dots\wedge p_m\in Y\wedge Y\neq t$ from which we
have obtained
$\{F_Y,\sigma_C^*(p_1),\dots,\sigma_C^*(p_m)\,|\,M_Y\}\neq
\sigma_C^*(t)$, then we had, from the definition of solved form,
that $Y$ does not occur in $t$, hence $F_Y$ does not occur at
depth $1$ in $\sigma_C^*(t)$, hence we do not obtain a disequation
of the form $n_{F_Y}\neq n_{F_Y}+1-1$.

{F}rom the safeness property, it is possible to find an integer
solution to the system $S$ by choosing arbitrarily large values
satisfying the constraints. Let $\{n_1 = \bar{n}_1,\dots, n_j =
\bar{n}_j\}$ be a solution and define $$\theta_2 = [R_i / \{ \nil
\}^ {\bar{n}_i}: \forall i \in \{1,\dots, j\}] \,.$$ where $ \{
\nil \}^ {\bar{n}}$ denotes the term
$\underbrace{\{\cdots\{}_{\bar{n}} \nil \}\cdots\}$ (similarly
for the other theories employed).

Let $\gamma = \theta_1 \sigma_C^*\theta_2$ (where $s \mu\nu$
stands for $(s\mu)\nu$) and observe that $C\gamma$ is a
conjunction of ground literals. We show that $KE^s_kF^s_1F_2F^3_s
\models C\gamma$. We analyze each literal of $C$\/.

\begin{description}

\item [$X = t:$] $\theta_1(X)$ coincides syntactically with
    $\theta_1(t)=t$\/. Hence, a literal of this form is true
in any model of equality.

\item [$t \in X:$]
$\theta_2(\sigma_C^*(X))=\{\dots,\theta_2(\sigma_{C}^*(t)),
\dots\}$\/,
so the atom is satisfied.

\item [$Z\neq u:$] two cases are possible:
   \begin{enumerate}
    \item if there are no atoms of the form $t\in Z$ in $C$\/,
          then the conditions in $S$ and over $\bar{s}$ ensure
          that ${\it rank}(\gamma(Z))\neq {\it rank}(\gamma(u))$\/;
    \item if there is at least one atom of the form $t\in Z$ in $C$\/,
          then $\sigma_C^*(Z)=\{F, t_{1},\dots,t_{k}\,|\,M\}$\/,
          the conditions in $S$ and over $\bar{s}$ ensure
          that ${\it rank}(\gamma(F))\neq {\it rank}(\gamma(u))-1$,
          hence $\gamma(F)$ is
          not an element of $\gamma(u)$.
    \end{enumerate}

\item  [$r \notin Y:$] two cases are possible:
    \begin{enumerate}
    \item no atoms of the form $t\in Y$ occur in $C$\/:
          if $r$ is ground, then it can not be an element of $Y$
          since $\gamma(Y)=\{\nil\}^{i}$, with $i\geq \bar{s}$\/; if $r$
          is not ground, then the conditions in $S$ ensure that ${\it
          rank}(\gamma(Y))\neq {\it rank}(\gamma(r))+1$\/;
    \item at least one atom of the form $t\in Y$ occurs in $C$\/, hence
          $\sigma_{C}^*(Y)=\{F,t_{1},\dots,t_{k}\,|\,M\}$\/: if $r$
          is ground the result is trivial; if $r$ is not ground then the
          conditions in $S$ ensure that
          ${\it rank}(\gamma(t_j))\neq {\it rank}(\gamma(r))$ for
          all $j\leq k$,
          ${\it rank}(\gamma(F))\neq {\it rank}(\gamma(r))$,
          and
          ${\it rank}(\gamma(M))\neq {\it rank}(\gamma(r))+1$\/.
    \end{enumerate}
\end{description}
\end{proof}

\begin{remark}\label{refining}
The task of testing whether a pre-solved form constraint $C$ is in
solved form could be avoided in the cases of multisets
and sets, where all membership atoms can be removed. As a
matter of fact, in the privileged models considered for sets and
multisets it holds that:
 $$s \in t \leftrightarrow \exists N(t = \{s\,|\,N\}).$$
We can therefore replace each membership atom $s \in t$ with an
equi-satisfiable equality atom $t = \{s\,|\,N\}$ with $N$ a new
variable. This implies that the additional conditions on the
pre-solved form are not required at all, since membership atoms can
be removed.
\end{remark}

\section{Constraint Rewriting Procedures}
\label{constraint-rewriting}

In this section we describe the procedures that can be used to
rewrite a given constraint $C$ into a equi-satisfiable disjunction
of constraints in pre-solved form.
All the procedures have the same overall structure
shown in Figure~\ref{main loop}: they take a constraint $C$\/ as
their input and repeatedly select an conjunct $c$ in $C$
not in pre-solved form (if any) and apply one of the rewriting
rules to it. The procedure stops when the constraint $C$ is
in pre-solved or \false\ is a conjunct of the constraint.
The procedure is non-deterministic. Some rewriting rules
have two or more possible non-deterministic choices.
Each non deterministic computation returns a constraint of the form
above. However there is globally a finite set
$C_1,\dots,C_k$ of constraints non-deterministically
returned.
The input constraint $C$ and
the disjunction
$C_1 \vee \cdots \vee C_k$ are equi-satisfiable.

\begin{figure}[htb]
{\small
$$
\begin{array}{||c|rcl||}
\multicolumn{4}{l}{\mbox{\sf \emph{Let $\Th$ be one of the
theories \lst, \bag, \clist, \set, $\pi$ a symbol in
\{$=$,$\neq$,$\in$,$\not\in$\}, and $C$ a $\Th$-constraint}}}\\
\multicolumn{4}{l}{\vspace{-5pt}}\\
\hline\hline \multicolumn{4}{||l||} {\phantom{aaa}\mbox{\sf while
$C$ contains
an atomic constraint $c$ of the form $\ell \pi r$ not in pre-solved
form and $c \neq \false$ do}}\\
\multicolumn{4}{||l||}
{\phantom{aaaaaa}\mbox{\sf select $c$\/;}}\\
\multicolumn{4}{||l||}
 {\phantom{aaaaaa}\mbox{\sf if $c$ = \false\ then return \false}}\\
\multicolumn{4}{||l||}
 {\phantom{aaaaaa}\mbox{\sf else if $c$ = \true\ then erase $c$}}\\
\multicolumn{4}{||l||}
 {\phantom{aaaaaa}\mbox{\sf else apply to $c$ any
rewriting rule for $\Th$-constraints of the form $\cdot \pi \cdot$\/;} }\\
\multicolumn{4}{||l||}
{\phantom{aaa}\mbox{\sf return $C$}}\\
\hline\hline
\end{array} $$}
\vspace{-5pt}\caption{\label{main loop}Main loop of constraint
rewriting procedures}
\end{figure}

\subsection{Equality Constraints} \label{sec.unif}

Unification algorithms for verifying the satisfiability and
producing the solutions of equality constraints in the four
aggregate's theories  have been proposed
in~\cite{DPR98-fuin}. The
unification algorithms proposed in~\cite{DPR98-fuin}
fall in the general schema of Figure~\ref{main loop}.
Some determinism in the statement {\sf select $c$} is
added to ensure termination.
They are called:
\begin{center}\begin{tabular}{lcl}
{\sf Unify\_lists} for lists & \phantom{aaaa}&
{\sf Unify\_msets} ({\sf Unify\_bags} in~\cite{DPR98-fuin})
   for multisets\\
{\sf Unify\_clists} for compact lists &&
{\sf Unify\_sets} for sets
\end{tabular}\end{center}
and they are used unaltered in the four global constraint
solvers  that we propose in this paper.

The output of the algorithms is either $\false$\/, when the
constraint is unsatisfiable, or a collection of solved form
constraints (Def.~\ref{solvedform}) composed only by equality
atoms. In Figure \ref{algobags.a} we have reported the rewriting
rules for the multisets unification used in algorithm
{\sf Unify\_msets}.

\begin{figure}[htb]
\small
$$\begin{array}{||c|rcl||}
\hline\hline
\multicolumn{4}{||l||}{\mbox{\sf \emph{Rules for {\sf Unify\_msets}}}}\\
\hline
  (1) &
        X =  X \phantom{iii}  &
        \mapsto & \true \\
  \hline
  (2) &
        \left.\begin{array}{r }
          t =  X \\
          \mbox{\small $t$ is not a variable}
        \end{array}\right\} &
        \mapsto  &  X =  t  \\
  \hline
  (3) &
        \left.\begin{array}{r}
             X =  t \\
             \mbox{\small $X$ does not occur in $t$, $X$ occurs in $C$}
        \end{array}\right\} &
        \mapsto  & \\
   &  \multicolumn{3}{r||}{X = t \mbox{ {\small and apply the
       substitution $X/t$ to $C$}}}\\
    \hline
    (4) &
        \left.\begin{array}{r}
             X =  t  \\
             \mbox{\small $X$ is not $t$ and $X$ occurs in $t$}
        \end{array}\right\} &
        \mapsto  & \false \\
  \hline
  (5) &
        \left.\begin{array}{r}
        f(s_1,\dots,s_m) =
        g(t_1,\dots,t_n) \\
        f \mbox{ is not } g
        \end{array}\right\} &
        \mapsto & \false \\
  \hline
  (6) &
         \left.\begin{array}{r}
                      f(s_1,\dots,s_m) =
              f(t_1,\dots,t_m)   \\
                        m \geq 0,  f  \mbox{ is not } \mf
                \end{array}\right\} &
        \mapsto &   \\
                &  \multicolumn{3}{r||}{s_1 =  t_1 \wedge \dots
        \wedge s_m =  t_m} \\
  \hline
  (7) &\left.
          \begin{array}{r}
            \mo t\,|\,s\mc =  \mo t'\,|\,s'\mc \\
            \mbox{\small ${\sf tail}(s)$ and ${\sf tail}(s')$
            are not the same variable}
          \end{array}
        \right\} & \mapsto &\\
        & \multicolumn{3}{r||}{
         \begin{array}{cl}
        (i) &  (t =  t' \wedge s =  s') \vee\\
        (ii) & (s =  \mo t'\,|\,N\mc \wedge
                 \mo t\,|\,N \mc =  s')
                 \end{array}} \\
  \hline
  (8) &\left.
          \begin{array}{r}
            \mo t\,|\,s\mc =  \mo t'\,|\,s'\mc \\
            \mbox{\small ${\sf tail}(s)$ and  ${\sf tail}(s')$
            are the same variable}
          \end{array}
                \right\} & \mapsto & \\
      & \multicolumn{3}{r||}{
               {\sf untail}(\mo t\,|\,s\mc)  =
               {\sf untail}(\mo t'\,|\,s'\mc )}\\
\hline\hline
\end{array}$$
\caption{\label{algobags.a}Rewriting rules for the Unification
algorithm for multisets}
\end{figure}

\label{untail}
The algorithm uses the auxiliary
functions ${\sf tail}$ and ${\sf untail}$ defined
as follows:
$$\begin{array}{lcll}
  {\sf tail}(f(t_1,\dots,t_n)) & = & f(t_1,\dots,t_n) &
   \mbox{$f$ is not $\mf$, $n\geq 0$}\\
  {\sf tail}(X) & = & X & \mbox{$X$ is a variable}\\
  {\sf tail}(\mo t\,|\, s \mc) & =  &
  {\sf tail}( s )\\
  {\sf untail}(X) & =  & \nil& \mbox{$X$ is a variable}\\
  {\sf untail}(\mo t\,|\, s \mc) & = &
           \mo t\,|\, {\sf untail}( s ) \mc
 \end{array}$$

\subsection{Membership and not-Membership Constraints}

The rewriting rules for membership and not-membership constraints
are justified by axioms $(K)$ and $(W)$ that hold in all the four
theories. Therefore, in Figure~\ref{innonin} we  give a single
definition of these rules. They are used within the main loop in
Figure~\ref{main loop} to define the rewriting procedures for
membership and not-membership constraints over the considered
aggregate. When useful, we will refer to these procedures with the
generic names {\sf in-$\Th$} and {\sf nin-$\Th$}, where $\Th$ is
any of the aggregate theories.

\begin{figure}[htb]
{\small
$$\begin{array}{||c|rcll||}
\multicolumn{5}{l}{\mbox{\sf \emph{Let $\cons$ be the aggregate
constructor for the theory $\Th$}}}\\
\hline\hline
\multicolumn{5}{||l||}{\mbox{\sf \emph{Rules for {\bf in-$\Th$}}}}\\
\hline
(1) &   \left.\begin{array}{r}
 r \in f(t_1,\dots,t_n)\\
                   f \mbox{ is not } \cons
  \end{array}\right\} & \mapsto & {\false}  & \\
\hline
(2) &   \left.\begin{array}{r} r \in {\tt cons_\Th}(t,s)
 \end{array}\right\}  & \mapsto & r = t \:\vee & (a) \\
   &&& r \in s & (b)\\
\hline
(3) & \left.\begin{array}{r}
          r \in X \\
        X \in \vars(r)
        \end{array}\right\} & \mapsto & {\false} & \\
\hline\hline
\end{array}$$

$$\begin{array}{||c|rcl||}
\hline\hline
\multicolumn{4}{||l||}{\mbox{\sf \emph{Rules for {\bf nin-$\Th$}}}}\\
\hline
(1) &   \left.\begin{array}{r}
           r \notin f(t_1,\dots,t_n)\\
                   f \mbox{ is not } \cons
        \end{array}\right\} & \mapsto & {\true} \\
\hline
(2) & \left.\begin{array}{r}
          r \notin {\tt cons_\Th}(t,s)
      \end{array}\right\} & \mapsto &
      r \neq t \wedge r \notin s \\
\hline
(3) & \left.\begin{array}{r}
          r \notin X \\
        X \in \vars(r)
        \end{array}\right\} & \mapsto & {\true} \\
\hline \hline
\end{array}$$}

\vspace{-5pt}\caption{\label{innonin}Parametric rewriting rules
for membership and not-membership constraints}\vspace{+10pt}
\end{figure}

\begin{lemma}\label{tuttimember}
Let $\Th$ be one of the theories \lst, \bag, \clist, \set, and
${\cal A}_{\Th}$ the privileged model for the theory $\Th$\/.
Let $C$ be a $\Th$-constraint, $C_1,\dots,C_k$ be the constraints
non-deterministically returned by {\sf
nin-{$\Th$}(in-{$\Th$}$(C)$))}\/, and $\bar N_i = \vars(C_i)
\setminus \vars(C)$\/. Then ${\cal A}_{\Th} \models \forall
\left(C \leftrightarrow \bigvee_{i=1}^{k} \exists \bar N_i C_i
\right)$\/.
\end{lemma}
\begin{proof}
We prove correctness and completeness for lists, thus with respect
to the model $\cal LIST$\/. Soundness and completeness for the
other aggregates are proved in the very same way. Soundness and
completeness is proved for each rewriting rule separately since
the rules are mutually exclusive. When possible, we simply point
out the axioms of the corresponding theory \lst\ involved in the
proof (note that $\cal LIST$ is a model of those axioms):
\begin{description}
\item[\mbox{{\sf in-List}, {\rm rule} $(1)$}\/.] $r \in
f(t_1,\dots,t_n)$\/, with $f$ different from $[ \cdot \,|\, \cdot ]$ is
equivalent to {\true}\ by axiom $(K)$\/.
\item[\mbox{{\sf in-List}, {\rm rule} $(2)$}\/.] This is
exactly axiom $(W)$\/.
\item[\mbox{{\sf in-List}, {\rm rule} $(3)$}\/.]
Assume that there is a valuation
$\sigma$ such that ${\cal LIST} \models \sigma(r \in X)$\/.
   This means that $\sigma(X)$ contains a term of the form:
   $[ s_1 , \dots, s_n , r'
   \,|\, t]$ for some terms $s_1,\dots,s_n,t$\/,
   and some term $r'$ in $\sigma(r)$. Axiom $(F_3)$
   ensures that $X$ can not be a subterm of $r$\/.
\item[\mbox{{\sf nin-List}, {\rm rules} $(1)$\/, $(2)$\/, $(3)$}\/.]
Same proofs as for the corresponding {\sf in-List} rules, using
the same axioms.
\end{description}
\end{proof}

In the above lemma it holds that the lists of variables
$\bar N_i$ are all empty.
However, for the sake of uniformity with respect to the
other similar correctness results, we have made them explicit.
Let us observe that the rewriting rules for procedure {\sf
in-MSet} and {\sf in-Set} could safely be extended by the rule:
{\small
$$\renewcommand{\arraystretch}{\scala}
\begin{array}{||c|rcl||}
\hline\hline (4) &  \phantom{aaa}   \left.\begin{array}{r}
          r \in X \\
        X \not\in \vars(r)
        \end{array}\right\} & \mapsto &
X = \mo r \,|\, N \mc \phantom{aaa}(X = \{ r \,|\, N \}) \\
\hline\hline
\end{array}$$
}

\noindent where $N$ is a new variable (see also
Remark~\ref{refining}). In this way, we are sure to completely
remove membership atoms from the constraints and that the
pre-solved form constraints obtained are in solved form.

\subsection{Disequality constraints}

Rewriting rules for disequality constraints consist of a part
which is the same for the four theories (although parametric with
respect to the considered theory), and a part which is specific
for each one of the four theories. Rules of the common part are
shown in Figure~\ref{algo-neq-general}, while specific rules are
described in the next subsections.

\begin{figure}[htb]
{\small
$$\begin{array}{||c|rcl||}
\multicolumn{4}{l}{\mbox{\sf \emph{Let $\cons$ be the aggregate
constructor for the theory $\Th$}}}\\
\multicolumn{4}{l}{\vspace{-5pt}}\\
\hline \hline
\multicolumn{4}{||l||}{\mbox{\sf \emph{Rules for {\bf neq-$\Th$}}}}\\
\hline
 (1) &  \left.\begin{array}{r}
             d \neq d \\
                          d \mbox{ is a constant}
           \end{array}\right\}
                         & \mapsto & {\false} \\
\hline
 (2) & \left.\begin{array}{r}
             f(s_1,\dots,s_m) \neq g(t_1,\dots,t_n) \\
                         f \mbox{ is not } g
          \end{array}\right\} &
        \mapsto & {\true} \\
\hline
 (3) & \left.\begin{array}{r}
             t \neq X \\
        \mbox{$t$ is not a variable}
      \end{array}\right\} & \mapsto & X \neq t \\
\hline
 (4) &  \left.\begin{array}{r}
             X \neq X  \\
                          X \mbox{ is a variable}
           \end{array}\right\}
                         & \mapsto & {\false} \\
\hline
 (5) & \left.\begin{array}{r}
         f(s_{1},\dots,s_n)
                 \neq
                 f(t_{1},\dots,t_n)  \\
                 n > 0, f \mbox{ is not } \cons
      \end{array}\right\}  & \mapsto &
          \begin{array}[t]{lc}
            s_{1} \neq t_{1} \vee & (1)\\
                \vdots & \vdots\\
       s_{n} \neq t_{n} & (n)
          \end{array} \\
\hline\hline
\end{array} $$}
\vspace{-5pt}\caption{\label{algo-neq-general}General rewriting
rules for disequality constraints} \vspace{+10pt}
\end{figure}

\subsubsection{Lists}

Specific rules for the theory \lst\ are presented in Figure
\ref{algo-neq-list}. These rules are inserted in the general
schema of Figure~\ref{main loop} to generate the procedure
\mbox{\sf  neq-List}.

\begin{figure}[htb]
{\small
$$\begin{array}{||c|rcl||}
\hline\hline
\multicolumn{4}{||l||}{\mbox{\sf \emph{Rules for {\bf neq-List}}}}\\
\hline
$(1)$-$(5)$ & {\mbox{\sf see Figure~\ref{algo-neq-general}}} & &\\
\hline
 (6) & \left.\begin{array}{r}
         [ s_{1}\,|\, s_2 ]
                 \neq
         [ t_{1}\,|\,t_2 ]
      \end{array}\right\}  & \mapsto &
          \begin{array}[t]{lc}
            s_{1} \neq t_{1} \vee & (i)\\
            s_{2} \neq t_{2} & (ii)
          \end{array} \\
\hline
 (7) & \left.\begin{array}{r}
          X \neq f(t_{1},\dots,t_n) \\
          X \in \vars(t_{1},\dots,t_n)
      \end{array}\right\}
          & \mapsto & {\true} \\
\hline\hline
\end{array} $$}
\vspace{-5pt}\caption{\label{algo-neq-list}Rewriting rules for
disequality constraints over lists} \vspace{+10pt}
\end{figure}

\begin{lemma}\label{corrlist}
Let $C$ be a \lst-constraint, $C_1,\dots,C_k$ be the constraints
non-deterministically returned by {\sf neq-List}$(C)$\/, and $\bar
N_i = \vars(C_i) \setminus \vars(C)$\/. Then $\mbox{\lst} \models
\forall \left(C \leftrightarrow \bigvee_{i=1}^{k} \exists \bar
N_i C_i \right)$\/.
\end{lemma}
\begin{proof}
Soundness and completeness of the rewriting rules (and, hence, of
the whole rewriting procedure {\sf neq-List}) are immediate
consequence of standard equality axioms and axiom schemata
$(F_1),(F_2),$ and $(F_3)$\/.
\end{proof}

\subsubsection{Multisets}

Disequality constraints over multisets are simplified using the
rewriting rules presented in Figure~\ref{neq-bag-fig}. They make
use of functions ${\sf tail}$ and ${\sf untail}$ defined in
Section~\ref{untail}. Using these rules within the generic
rewriting scheme of Figure~\ref{main loop} we get the rewriting
procedure for disequality constraints over multisets, called {\sf
neq-MSet}.
\begin{figure}[htb]
{\small
$$\begin{array}{||c|rcl||}
\hline\hline
\multicolumn{4}{||l||}{\mbox{\sf \emph{Rules for {\bf neq-MSet}}}}\\
\hline
 $(1)$-$(5)$ & {\mbox{\sf see Figure~\ref{algo-neq-general}}} & &\\
\hline
 (6.1) & \left.\begin{array} {r}
        \mo t_1 \,|\, s_1\mc \neq \mo t_2 \,|\, s_2 \mc\\
        \mbox{${\sf tail}(s_1)$ and ${\sf tail}(s_2)$}\\
        \mbox{are the same variable}
      \end{array}\right\} & \mapsto &
       {\sf untail}(\mo t_1 \,|\, s_1 \mc) \neq
       {\sf untail}(\mo t_2 \,|\, s_2 \mc)\\
\hline
(6.2) & \left. \begin{array} {r}
      \mo t_1 \,|\, s_1 \mc \neq \mo t_2 \,|\, s_2 \mc\\
      \mbox{${\sf tail}(s_1)$ and ${\sf tail}(s_2)$}\\
        \mbox{are not the same variable}
      \end{array} \right\} & \mapsto &
      \begin{array}[t]{lc}
          (t_1 \neq t_2 \wedge t_1 \notin s_2) \vee & (a)\\
      (\mo t_2 \,|\, s_2 \mc = \mo t_1 \,|\, N \mc
          \wedge s_1 \neq  N ) & (b)
          \end{array}\\
\hline
 (7) & \left.\begin{array}{r}
          X \neq f(t_{1},\dots,t_n) \\
          X \in \vars(t_{1},\dots,t_n)
      \end{array}\right\}
          & \mapsto & {\true} \\
\hline\hline
\end{array}$$}
\vspace{-5pt}\caption{\label{neq-bag-fig} Rewriting rules for
disequality constraints over multisets}\vspace{+10pt}
\end{figure}

Some words are needed for explaining the rules related to the
management of disequalities between multisets; in particular rule
$(6.2)$ of Figure~\ref{neq-bag-fig}. If we use directly axiom
$(E^m_k)$\/, we have that:
$$\begin{array}{rcl}
\mo t_1 \,|\, s_1 \mc \neq \mo t_2 \,|\, s_2 \mc &
\leftrightarrow  & (t_1 \neq t_2 \vee s_1 \neq s_2) \wedge\\
&& \forall N\,(s_2 \neq \mo t_2 \,|\, N \mc \vee
            s_1 \neq \mo t_1 \,|\, N \mc)
\end{array}$$
This way, an universal quantification is introduced: this is
no longer a constraint according to Definition~\ref{constraintdef}.

Alternatively, we could use the intuitive notion of
multi-membership: $x \in^i y$ if $x$ belongs at least $i$ times to the
multiset $y$\/. This way, one can provide an alternative version
of equality and disequality between multisets. In particular, we
have:
$$\begin{array}{rcl}
\mo t_1 \,|\, s_1 \mc \neq \mo t_2 \,|\, s_2 \mc &
\leftrightarrow  & \exists X \exists n \,( n \in \nat
\,\wedge \\
&& \phantom{aaaaaa}
(X \in^n  \mo t_1 \,|\, s_1 \mc \wedge
X \notin^n  \mo t_2 \,|\, s_2 \mc) \vee\\
&& \phantom{aaaaaa}
(X \in^n   \mo t_2 \,|\, s_2 \mc \wedge
X \notin^n    \mo t_1 \,|\, s_1 \mc\mo t_2 \,|\, s_2 \mc))
\end{array}$$
In this case, however, we have a quantification on natural
numbers: we are outside the language we are studying.
The rewriting rule $(6.2)$ adopted in Figure~\ref{neq-bag-fig}
avoids these difficulties introducing only existential
quantification. Its correctness and completeness are proved in
the following lemma.

\begin{lemma}\label{corrbag}
Let $C$ be a \bag-constraint, $C_1,\dots,C_k$ be the constraints
non-deterministically returned by {\sf neq-MSet}$(C)$\/, and $\bar
N_i = \vars(C_i) \setminus \vars(C)$\/. Then ${\cal MSET} \models
\forall \left(C \leftrightarrow \bigvee_{i=1}^{k} \exists \bar
N_i C_i \right)$\/.
\end{lemma}
\begin{proof}
{F}rom Lemma~\ref{corrlist} we know that the result holds for
rules $(1)$--$(5)$ and $(7)$ for the model $\cal LIST$\/. Since
permutativity has not been used for that result, and axiom $(F_3)$
holds for both the theories, the same holds for the model $\cal
MSET$\/. We need to prove correctness and completeness of
rewriting rules $(6.1)$ and $(6.2)$\/.
\begin{description}
\item[$(6.1)$]   It is immediately justified by axiom schema $(F_3^m)$.

\item[$(6.2)$]  The constraint
   $\mo t_1\,|\,s_1  \mc \neq  \mo t_2\,|\,s_2  \mc$
is equivalent to:
\begin{eqnarray}
&& t_1 \notin \mo t_2\,|\,s_2  \mc \wedge \mo t_1\,|\,s_1
\mc
\neq  \mo t_2\,|\,s_2  \mc
\, \vee\label{formulaa} \\
&& t_1 \in \mo t_2\,|\,s_2  \mc \wedge \mo t_1\,|\,s_1  \mc
\neq  \mo t_2\,|\,s_2  \mc
\label{formulab}
\end{eqnarray}
Since we are looking for successful valuations over $\cal MSET$
that deal with multisets of finite elements, axiom $(E^m_k)$
ensures that $t_1 \notin \mo t_2\,|\,s_2 \mc $  implies
 $\mo t_1\,|\,s_1  \mc \neq  \mo t_2\,|\,s_2  \mc$\/.
 Thus, formula (\ref{formulaa}) is equivalent to
 $t_1 \in \mo t_2\,|\,s_2 \mc $ which, in turn, is equivalent
 by $(W)$ to the disjunct $(a)$ of the rewriting rule.

Consider now formula (\ref{formulab}).
It is easy  to see that
\begin{eqnarray}\label{formulac}
{\cal MSET} \models \forall (t_1 \in \mo t_2\,|\,s_2 \mc
\leftrightarrow \exists M \, (
       \mo  t_1 \,|\, M \mc =
        \mo t_2\,|\,s_2  \mc))
\end{eqnarray}
Thus, (\ref{formulab}) is equivalent
to
\begin{eqnarray}\label{formulad}
\exists M \, (
       \mo  t_1 \,|\, M \mc =
        \mo t_2\,|\,s_2  \mc
                \wedge \mo t_1\,|\,s_1  \mc \neq  \mo
t_2\,|\,s_2  \mc )
\end{eqnarray}
It remains to prove that (\ref{formulad}) is equivalent to
the disjunct $(b)$\/, namely:
\begin{eqnarray}\label{formulas}
\exists N\, ( s_1 \neq N \wedge
           \mo t_2 \,|\, s_2 \mc = \mo t_1 \,|\, N \mc )
\end{eqnarray}

\begin{description}
\item[\rm $(\ref{formulad}) \rightarrow (\ref{formulas})$]
  Assume there is $M$ so as to satisfy $(\ref{formulad})$\/.
   $M = s_1$ will immediately lead to a contradiction.
   Thus, $ (\ref{formulas})$ is satisfied by $N = M$\/.

\item[\rm $(\ref{formulas}) \rightarrow (\ref{formulad})$]
  Assume there is $N$ so as to satisfy  $(\ref{formulas})$\/.
It follows immediately from the fact, true for finite multisets,
that $s_1 \neq N$ implies $\mo t_1 \,|\, s_1 \mc \neq \mo t_1
\,|\, N \mc$\/. Thus, choose $M = N$\/.

\end{description}\end{description}
\end{proof}

\subsubsection{Compact Lists}

The rewriting rules for disequality constraints over compact lists
are shown in Figure~\ref{neq-clist-fig}.
These rules can be immediately exploited in conjunction with the
generic scheme of Figure~\ref{main loop} to obtain a rewriting
procedure for disequality constraints over multisets--called {\sf
neq-CList}. Soundness and completeness of {\sf neq-CList} are
stated by the following lemma.

\begin{figure}[htb]
{\small
$$\begin{array}{||c|rcl||}
\hline\hline
 \multicolumn{4}{||l||}{\mbox{\sf \emph{Rules for {\bf neq-CList}}}}\\
\hline
 $(1)$-$(5)$ & {\mbox{\sf see Figure~\ref{algo-neq-general}}} & &\\
\hline
 (6) &  \left.\begin{array}[t]{r}
      \clo t_1 \,|\, s_1 \clc \neq \clo t_2 \,|\, s_2 \clc
  \end{array}\right\}
 & \mapsto & \\
 & \multicolumn{3}{|r||}{
 \begin{array}[t]{lc}
        t_1 \neq t_2 \vee & (a)\\
       s_1 \neq s_2 \wedge
       \clo t_1\,|\,s_1 \clc \neq s_2 \wedge
       s_1 \neq \clo t_2 \,|\, s_2 \clc & (b)
  \end{array}} \\
\hline
 (7.1) &  \left.
    \begin{array}{r}
      X \neq f(t_1,\dots,t_n)  \\
          X \in \vars(t_1,\dots,t_n), f \mbox{ is not } \lcf\\
   \end{array}\right\}
          & \mapsto & {\true} \\
\hline
 (7.2) & \left.\begin{array}{r}
      X \neq \clo t_1,\dots,t_n \,|\,X\clc \\
          X \in \vars(t_1,\dots,t_n)
          \end{array}\right\}
          & \mapsto & {\true} \\
\hline
 (7.3) & \left.\begin{array}{r}
      X \neq \clo t_1,\dots,t_n \,|\,X\clc \\
          X \notin \vars(t_1,\dots,t_n)
          \end{array}\right\}
& \mapsto &
\begin{array}[t]{lc}
t_1 \neq t_2 \vee & (a.1)\\
\vdots & \vdots \\
 t_1 \neq t_n \vee & (a.n)\\
 X = \nil\; \vee & (b) \\
 X = \clo N_1 \,|\, N_2 \clc \wedge N_1 \neq t_1 & (c)
\end{array}\\
\hline\hline
\end{array}$$}
\vspace{-5pt}\caption{\label{neq-clist-fig} Rewriting rules for
disequality constraints over compact lists}\vspace{+10pt}
\end{figure}

\begin{lemma}\label{corrclists}
Let $C$ be a \clist-constraint, $C_1,\dots,C_k$ be the constraints
non-deterministically returned by {\sf neq-CList}$(C)$\/, and
$\bar N_i = \vars(C_i) \setminus \vars(C)$\/. Then ${\cal CLIST}
\models \forall \left(C \leftrightarrow \bigvee_{i=1}^{k}
\exists \bar N_i C_i \right)$\/.
\end{lemma}
\begin{proof}
For rules $(1)$--$(5)$ the result follows immediately from those for lists.
Rules $(7.1)$--$(7.3)$ follows from axiom $(F^c_3)$.
Rule $(6)$ is exactly axiom $(E^c_k)$\/.
\end{proof}

Observe that, differently from multisets, the rewriting rule for
disequality between compact lists follows immediately from axiom
$(E^c_k)$\/. As a matter of fact, this axiom does not introduce
any new variable.

\subsubsection{Sets}

Disequality constraints over sets are dealt with by the rewriting
rules shown in Figure~\ref{neqsetfig}, and they constitute the
procedure {\sf neq-Set}.

\begin{figure}[htb]
{\small
$$\begin{array}{||c|rcl||}
\hline\hline
 \multicolumn{4}{||l||}{\mbox{\sf \emph{Rules for {\bf neq-Set}}}}\\
\hline
 $(1)$-$(5)$ & {\mbox{\sf see Figure~\ref{algo-neq-general}}} & &\\
\hline (6) &  \left.\begin{array}[t]{r}
      \{ t_1 \,|\, s_1 \} \neq \{ t_2 \,|\, s_2 \}
  \end{array} \right\}
 & \mapsto & \\
 & \multicolumn{3}{|r||}{
 \begin{array}[t]{lc}
        Z \in \{ t_1 \,|\, s_1 \} \wedge
                Z \notin \{ t_2 \,|\, s_2 \} \vee & (a)\\
       Z \in \{ t_2 \,|\, s_2 \} \wedge
       Z \notin \{ t_1 \,|\, s_1 \} & (b)
  \end{array}} \\
\hline
 (7.1) &  \left.
      \begin{array}{r}
      X \neq f(t_1,\dots,t_n)  \\
          X \in \vars(t_1,\dots,t_n), f \mbox{ is not } \sef\\
   \end{array}\right\}
          & \mapsto & {\true} \\
\hline
 (7.2) & \left.
      \begin{array}{r}
      X \neq \{ t_1,\dots,t_n \,|\,X\} \\
          X \in \vars(t_1,\dots,t_n)
          \end{array}\right\}
          & \mapsto & {\true} \\
\hline
 (7.3) & \left.\begin{array}{r}
      X \neq \{ t_1,\dots,t_n \,|\,X\} \\
          X \notin \vars(t_1,\dots,t_n)
          \end{array}\right\}
& \mapsto &
\begin{array}[t]{lc}
t_1 \notin X \vee & (i)\\
\vdots & \vdots \\
 t_n \notin X   & (n)
\end{array}\\
\hline\hline
\end{array}$$}
\vspace{-5pt}\caption{\label{neqsetfig}Rewriting rules for
disequality constraints over sets}\vspace{+10pt}
\end{figure}

Some remarks are needed regarding rule $(6)$\/. As for multisets,
axiom $(E^s_k)$ introduces an existentially quantified variable to
state equality. Thus, its direct application for stating
disequality requires universally quantified constraints that go
outside the language. On the other hand, the rewriting rule
$(6.2)$ used for multisets can not be used in this context. In
fact, the property that $s_1 \neq N$ implies $\mo t_1 \,|\, s_1
\mc \neq \mo t_1 \,|\, N \mc$\/, that holds for finite multisets,
does not hold for sets. For instance, $\{ a\} \neq \{a,b\}$ but
$\{ b , a \} = \{ b,a,b\}$\/. Thus, this rewriting rule would be
not correct for sets.

A rewriting rule for disequality constraints over sets can be
obtained by taking the negation of the standard extensionality
axiom
$$
\renewcommand{\arraystretch}{\scala}
\begin{array}{|crcl|}
\hline
(E_k) & x = y & \leftrightarrow  &
\forall z \, (z \in x \leftrightarrow z \in y) \\
\hline
\end{array}$$

\begin{lemma}\label{corrset}
Let $C$ be a \set-constraint, $C_1,\dots,C_k$ be the constraints
non-determini\-stically returned by {\sf neq-Set}$(C)$\/, and
$\bar N_i = \vars(C_i) \setminus \vars(C)$\/. Then ${\cal SET}
\models \forall \left(C \leftarrow \bigvee_{i=1}^{k} \exists
\bar N_i C_i \right)$\/.
\end{lemma}
\begin{proof}
For rules $(1)$--$(5)$ the result follows from those for lists.
Rules $(7.1)$ and $(7.2)$ are exactly axiom $(F^s_3)$\/.
Rule $(6)$ is axiom $(E_k)$\/,  implied by $(E^s_k)$ on
$\cal SET$.
\end{proof}

\begin{remark}

In our theories an aggregate can be built by starting from any
ground uninterpreted Herbrand term---called the
\emph{kernel}---and then adding to this term the elements
that compose the aggregate. Thus, two aggregates can contain the
same elements but nevertheless they can be different because of
their different kernels. For example, the two terms $\{a\,|\,b\}
\mbox{ and } \{a\,|\,c\}$ denote two different sets containing
the same elements ($a$) but based on different kernels ($b$ and
$c$, respectively).

Rewriting rules for disequality constraints over aggregates other
than sets are formulated in such a way to take care of the
(possibly different) kernels in the two aggregates without having
to explicitly resort to kernels. Conversely, the rewriting rule
for disequality constraints over sets (rule (6)) is not able to
``force'' disequality between two sets when they have the same
elements but different kernels. This the reason why the
$(\rightarrow)$ direction of Lemma \ref{corrset} does not hold.

A possible completion of the above procedures to take care of this
case is presented in~\cite{DR93}; for doing that some technical
complications are introduced. Basically, a new constraint
($\ker$) is introduced and the rewriting rule (6) is endowed with
a third non-deterministic case: $\ker(s_1) \neq \ker(s_2)$. The
advantage of this solution is completeness (the $(\rightarrow)$
direction of Lemma~\ref{corrset}). However, for the sake of
simplicity, we do not add here the details on the modifications of
the rewriting rules for dealing with $\ker$ that are instead
presented in~\cite{DR93}.
\end{remark}

\section{Constraint solving}\label{constraint-solving}

In this section we address the problem of establishing if a
constraint $C$ is satisfiable or not in the corresponding privileged
model. The correspondence result (Theorem \ref{corrispondenza})
ensures that the property is inherited by any model.

Constraint satisfiability for the theory $\Th$ is checked by the
non-deterministic rewriting procedure ${\sf SAT}_{\Th}$\/ shown in
Figure~\ref{figuraSAT}. Its definition is completely parametric
with respect to the theory involved. ${\sf SAT}_{\Th}$  uses
iteratively the various rewriting procedures presented in the
previous section, until a fixed-point is reached---i.e., any new
rewritings do not further simplify the constraint. This happens
exactly when the constraint is in pre-solved form or it is
\false. The two conditions that guarantee that a constraint in
pre-solved form is in solved form are tested by function ${\sf
is\_solved}_{\Th}$ shown in Figure~\ref{acyclic-proc}.

By Theorem~\ref{all solved} a constraint in solved form is
guaranteed to be satisfiable in the corresponding model.
Moreover, it will be proved (see Theorem~\ref{bagscorrect}) that
the disjunction of solved form constraints returned by ${\sf
SAT}_{\Th}$ is equi-satisfiable in that model to the original
constraint $C$\/. Therefore, ${\sf SAT}_{\Th}$ can be used as a
test procedure to check satisfiability of $C$: if it is able to
reduce $C$ to at least one solved form constraint $C'$ then $C$ is
satisfiable; otherwise, $C$ is unsatisfiable. Moreover, the
generated constraint in solved form can be immediately exploited
to compute all possible solutions for $C$\/.

\begin{figure}[htb]
{\small
$$\begin{array}{||rcl||}
\hline\hline
\multicolumn{3}{||l||}
{\mbox{\sf function {\sf SAT}$_{\Th}$}(C)}\\
\multicolumn{3}{||l||}
{\phantom{aaa}{\sf repeat}}\\
  & & \phantom{aaa} C' := C;\\
  & & \phantom{aaa} C := \mbox{\sf Unify\_{$\Th$s}}(
           \mbox{\sf neq-$\Th$}(
           \mbox{\sf nin-$\Th$}(
           \mbox{\sf in-$\Th$}(C))))\\
\multicolumn{3}{||l||}{\phantom{aaa} {\sf until}\;C = C';}\\
\multicolumn{3}{||l||}{\phantom{aaa} \mbox{\sf return(
      is\_solved}_{\Th}(C)).}\\
\hline\hline
\end{array}$$}
\vspace{-5pt}\caption{\label{figuraSAT}The satisfiability
procedure, parametric with respect to $\Th$}\vspace{+10pt}
\end{figure}

\begin{figure}
{\small
$$\begin{array}{||l||}
\hline\hline
\mbox{\sf function is\_solved$_{\Th}$}(C)\\
\phantom{aaa}\mbox{\sf build the directed graph ${\cal
G}_{C}^{\in}$};\\
\phantom{aaa}\mbox{\sf if ${\cal G}_{C}^{\in}$ has a
cycle}\\
\phantom{aaaaaa}\mbox{\sf then return \false}\\
\phantom{aaaaaa}\mbox{\sf else}\\
\phantom{aaaaaaaaa}\mbox{compute $\sigma_C^*$}\\
\phantom{aaaaaaaaa}\mbox{\sf if there is a pair $t \in X , t'
\notin X$
in $C$ s.t. $\Th \models \forall (\sigma_C^*(t) = \sigma_C^*(t'))$}\\
\phantom{aaaaaaaaaaaa}\mbox{\sf then return \false}\\
\phantom{aaaaaaaaaaaa}\mbox{\sf else return $C$\/.}\\
\hline\hline
\end{array}$$}
\vspace{-5pt}\caption{\label{acyclic-proc}Final check for solved
form constraints}\vspace{+10pt}
\end{figure}

The rest of this section is devoted to prove the crucial result of
termination for procedure ${\sf SAT}_{\Th}(C)$ and, then, to prove
its soundness and completeness.

\begin{theorem}[Termination]
\label{termina-lists} \label{termina-bags} \label{termina-clist}
\label{termina-set}

Let $\Th$ be one of the theories \lst, \bag, \clist, \set, and $C$
be a $\Th$-constraint. Each non-determi\-nistic execution of $ {\sf
SAT}_{\Th}(C) $\/ terminates in a finite number of steps.
Moreover, the constraint returned is either $\false$ or a solved
form constraint.
\end{theorem}
\begin{proof}
We give the proof for the case of \bag. The other proofs are in
Appendix~\ref{termination}.

It is immediate to see, by the definition of the procedures, that
if $C$ is different from $\false$ and not in pre-solved form,
then some rewriting rule can be applied. The function ${\sf
is\_solved_{\bag}}$, whose termination follows from termination
of {\sf Unify\_msets}~\cite{DPR98-fuin}, needed for the solved
form test $\Th \models\forall (\sigma_C^*(t) =
\sigma_C^*(t'))$\/, produces by definition solved form
constraints or \false.

We prove that the {\sf repeat} cycle can not loop forever. For
doing that, we define a complexity measure for constraints. Let us
assume that constraints of the form $X=t$\/, with $X$ neither in
$t$ nor elsewhere in $C$\/, are removed from $C$\/. Similarly, we
assume that $\true$ constraints are not counted in the complexity
measure. These two assumptions are safe since those constraints
do not fire any new rule application. The complexity measure that
we associate with a constraint is the following triple:
$$\begin{array}{rclll} compl(C) & = &  \langle &
\alpha(C) = \mbox{\# vars in $C$},\\ &&&  \beta(C) = \mo \size(s) +
\size(t) : s \mathbin{op} t \in C \mc , \\ &&&  \gamma(C) =
\sum_{s\:\mathbin{op}\:t \in C} \size(s) &\rangle
\end{array}$$

The first and third element of the tuple are non-negative
integers. The second is a multiset of non-negative integers.
They are well-ordered~\cite{DM79} by the ordering obtained as
the transitive closure of the rule:
$$\mo s_1,\dots,s_{i-1},t_1,\dots,t_n,s_{i+1},\dots,s_m\mc
  \prec
  \mo s_1,\dots, s_m\mc \,,$$
for $i \in \{1,\dots,m \}$\/, $n \geq 0$\/,
$t_{1} < s_{i},\dots,t_{n}< s_{i}$\/.
The ordering on triples
is the (well-founded) lexicographical ordering.

We will prove that given a constraint $C$, in a finite number of
non-failing successive rule applications,  a constraint $C'$ with
lower complexity is reached.  We show, by case analysis, this
property. Most rule applications decreases the complexity in one
step. When this does not happen, we enter in more detail.

\begin{description}
\item[{\sf Unify\_msets}$(1)$] $\alpha$ does not increase,
$\beta$ decreases.
\item[{\sf Unify\_msets}$(2)$] $\alpha$ and  $\beta$ do not increase.
$\gamma$ decreases, since $\size(X) = 0$ and $\size(t) > 0$\/.
\item[{\sf Unify\_msets}$(3)$] $\alpha$ decreases by 1.
\item[{\sf Unify\_msets}$(6)$] $\alpha$ does not increase.
$\beta$ decreases, since
  an equation of size $1 + \sum_{i=1}^m \size(s_i) +
\size(t_i)$
  is replaced by $m$ smaller equations of size $\size(s_i) +
  \size(t_i)$\/.
\item[{\sf Unify\_msets}$(7)$] In this case the complexity may
remain unchanged at the first step. However, the unification
algorithm adopts a selection strategy that ensures that after a
finite number of steps, we either reach a situation such that
$\alpha$ decreases or a situation where $\alpha$ is unchanged and
$\beta$ decreases (see~\cite{DPR98-fuin} for details).
\item[{\sf Unify\_msets}$(8)$] After one rule application, we
are in the case $(7)$ with both the tails of the multisets non
variables. After a finite number of steps, we enter the situation
where $\alpha$ is unchanged and $\beta$ decreases.
\item[{\sf in-MSet}$(2)$]
  $\alpha$ does not increase. $\beta$ decreases,  since
  a constraint of size $1 +  \size(r) + \size(s) + \size(t)$
  is non-deterministically replaced by one of smaller size
  $\size(r) + \size(s)$ or $\size(r) + \size(t)$\/.

\item[{\sf nin-MSet}$(1),(3)$]
  Trivially, $\alpha$ does not increase and $\beta$ decreases.

\item[{\sf nin-MSet}$(2)$]
  $\alpha$ does not increase. $\beta$ decreases,  since
  a constraint of size $1 +  \size(r) + \size(s) + \size(t)$
  is non-deterministically replaced by two of smaller size
  $\size(r) + \size(s)$ and $\size(r) + \size(t)$\/.
\item[{\sf neq-MSet}$(2),(7)$]
  Trivially, $\alpha$ does not increase and $\beta$ decreases.
\item[{\sf neq-MSet}$(3)$]
  $\alpha$ and  $\beta$ do not increase.
  $\gamma$ decreases, since $\size(X) = 0$ and $\size(t) > 0$.
\item[{\sf neq-MSet}$(5)$]
  $\alpha$ does not increase. $\beta$ decreases, since
  a constraint of size $1 + \sum_{i=1}^m \size(s_i) +
  \size(t_i)$
  is non-deterministically replaced by one of size
  $\size(s_i) + \size(t_i)$\/.
\item[{\sf neq-MSet}$(6.2)$] A unique application of
this rule may not decrease the constraint complexity.
Thus, we enter in some detail. The rule
removes
$\mo t_1 \,|\, s_1 \mc \neq \mo t_2 \,|\, s_2 \mc$
and introduces
\begin{eqnarray}
&&\mo t_2 \,|\, s_2 \mc = \mo t_1 \,|\, N \mc  \wedge
\label{stella}\\
&&s_1 \neq N \label{duestelle}
\end{eqnarray}
Consider now the two cases:
\begin{enumerate}
\item $\mo t_2 \,|\, s_2 \mc$ is $\mo r_{1},\dots,r_{n}\mc$
\item $\mo t_2 \,|\, s_2 \mc$ is $\mo
r_{1},\dots,r_{n}\,|\,A\mc$\/, for
some variable  $A$ distinct from $N$ that has just been
introduced.
\end{enumerate}
In the first case the  successive execution of {\sf Unify\_{bags}}
replaces equation (\ref{stella}) by:
$$t_{1} = r_{i}, N = \mo
r_{1},\dots,r_{i-1},r_{i+1},\dots,r_{n} \mc$$
for some $i = 1,\dots,n$\/.
We have that
$$size(t_{1}) + \size(r_{i}) <
 \size(\mo t_1 \,|\, s_1 \mc) + \size(\mo t_2 \,|\, s_2
\mc).$$ The equation $ N = \mo
r_{1},\dots,r_{i-1},r_{i+1},\dots,r_{n}\mc$ is eliminated by
applying the substitution for $N$\/. $N$ occurs only in the
constraint $s_{1} \neq N$\/, that becomes $s_{1} \neq \mo
r_{1},\dots,r_{i-1},r_{i+1},\dots,r_{n}\mc$\/. Again, its $size$
is strictly smaller than that of the original disequality
constraint. Thus, after some further steps, $\alpha$ remains
unchanged while $\beta$ decreases.
Strictly speaking, some other
actions may occur during that sequence of actions. However, if no
other rule $(6.2)$ is executed, then all rules decrease the
complexity tuples. Conversely, if other rules of this form are
executed, then we need to wait for all the substitutions of this
form to be applied. But they are all independent processes.

The second case is similar, but in this case a substitution
also for $A$ is computed, ensuring that $\alpha$ decreases.

\item[{\sf neq-MSet}$(6.1)$] After one step, we are in
the above situation $(6.2)$.
\end{description}
\end{proof}
The soundness and completeness result of the global constraint
solving procedure for \lst, \bag, and \clist\ follows from the
lemmas in the previous section and two lemmas in the
Appendix~\ref{modellistiche}.

\begin{theorem}[Soundness - Completeness]
\label{listcorrect} \label{bagscorrect} \label{clistcorrect}
\label{setcorrect} Let $\Th$ be one of the theories \lst, \bag,
\clist, and \set, $C$ be a $\Th$-constraint, and $C_1,\dots,C_k$
be the solved form constraints non-determi\-nistically returned by
${\sf SAT}_{\Th}(C)$\/, and $\bar N_i$ be $\vars(C_i) \setminus
\vars(C)$\/. Then ${\cal A}_{\Th} \models \forall \left(C
\leftrightarrow \bigvee_{i=1}^{k} \exists \bar N_i C_i \right)$\/,
where ${\cal A}_{\Th}$ is the model which corresponds with $\Th$.
\end{theorem}
\begin{proof}
Theorem~\ref{termina-lists} ensures the termination of each
non-deterministic branch. At each branch point, the number of
non-deterministic choices is finite. Thus, $C_1,\dots,C_k$ can be
effectively computed. Soundness and completeness follow from the
results proved individually for the procedures involved:
Lemma~\ref{tuttimember} for {\sf in}-$\Th$ and {\sf nin}-$\Th$\/;
Lemma~\ref{corrlist},  Lemma~\ref{corrbag},
Lemma~\ref{corrclists}, and Lemma~\ref{corrset} for {\sf
neq-MSet}, {\sf neq-List}, {\sf neq-CList}, and {\sf neq-Set},
respectively; Lemma~\ref{uffa} for ${\sf is\_solved_\Th(C)}$;
\cite{DPR98-fuin} for unification.
\end{proof}

\begin{corollary}[Decidability]\label{decidlist}
Given a $\Th$-constraint $C$\/, it is decidable whether ${\cal A}
\models   \exists\, C$\/, where ${\cal A}$ is one of the
privileged models $\cal LIST$\/, $\cal MSET$\/, $\cal CLIST$\/, $\cal
SET$\/.
\end{corollary}
\begin{proof}
{F}rom Theorem~\ref{listcorrect} we know that $C$ is
equi-satisfiable to $C_1 \vee \cdots \vee C_k$\/. If all the $C_i$
are $\false$\/, then $C$ is unsatisfiable in $\cal LIST$ ($\cal
MSET$\/, $\cal CLIST$\/, $\cal SET$). Otherwise, it is
satisfiable, since solved form constraints are satisfiable
(Theorem~\ref{all solved}).
\end{proof}

\subsection{Complexity Issues}

Complexity of the four unification problems is studied
in~\cite{DPR98-fuin}: the decision problem for unification
is proved to be solvable in linear time for lists, and
it is NP-complete for the other cases.

In the case of lists, if the constraint is a conjunction of
equality and disequality constraints, then the satisfiability
problem for a constraint $C$ is solvable in $O(n^{2})$ where $n =
|C|$~\cite{BN98,CB83}. Instead, the satisfiability problem for
conjunctions of membership and disequality constraints over lists
is NP-hard. As a matter of fact, let us consider the following
instance of 3-SAT:
$$(X_1 \vee X_2 \vee \neg X_3) \wedge
           (\neg X_1 \vee X_2 \vee X_3) \wedge
           (X_1 \vee \neg X_2 \vee X_3)\;.$$
The above instance of
3-SAT can be re-written as the following
constraint problem:
$$\begin{array}[t]{ccccc}
   X_1 \in [\underline{0}, \underline{1}] & \wedge &
   Y_1 \in [\underline{0}, \underline{1}] & \wedge & \\
   {[X_1,Y_1]} \neq [ \underline{0}, \underline{0}] & \wedge &
   [X_1,Y_1] \neq [ \underline{1}, \underline{1}] & \wedge &\\
   X_2 \in [\underline{0}, \underline{1}] & \wedge &
   Y_2 \in [\underline{0}, \underline{1}] & \wedge & \\
   {[X_2,Y_2]} \neq [ \underline{0}, \underline{0}] & \wedge &
    [X_2,Y_2] \neq [ \underline{1}, \underline{1}] & \wedge &\\
   X_3 \in [\underline{0}, \underline{1}] & \wedge &
   Y_3 \in [\underline{0}, \underline{1}] & \wedge & \\
   {[X_3,Y_3]} \neq [ \underline{0}, \underline{0}] & \wedge &
   [X_3,Y_3] \neq [ \underline{1}, \underline{1}] & \wedge &\\
   {[ X_1,X_2,Y_3]} \neq [ \underline{0}, \underline{0},
    \underline{0}] & \wedge &
   [ Y_1,X_2,X_3] \neq [ \underline{0}, \underline{0},
    \underline{0}] & \wedge &
   [ X_1, Y_2, X_3 ] \neq [ \underline{0}, \underline{0},
    \underline{0}]
  \end{array} $$
where $\underline{0}$ and $\underline{1}$ can be represented by
$\nil$ and $[\nil]$\/, respectively, and $Y_i$ takes the place of
$\neg X_i$ and vice versa. It is immediate to prove that any
substitution satisfying the constraint problem is also a solution
for the above formula, provided  $\underline{0}$ is interpreted as
{\false} and $\underline{1}$ is interpreted as {\true}, and vice
versa.

\section{Conclusions}\label{conclusions}

In this paper we have extended the results of~\cite{DPR98-fuin}
studying the constraint solving problem for four different
theories: the theories of lists, multisets, compact
lists, and sets. The analyzed constraints are conjunctions of
literals based on equality and membership predicate symbols. We
have identified the privileged models for these theories by
showing that they correspond with the theories on the class of
considered constraints. We have developed a notion of solved form
(proved to be satisfiable) and presented the rewriting algorithms
which allow this notion to be used to decide the satisfiability
problems in the four contexts.

In particular, we have shown how constraint solving can be
developed parametrically for these theories and we have pointed
out the differences and similarities between the four kinds of
aggregates.

As further work it could be interesting to study the properties of
the four aggregates in presence of
append-like operators (\emph{append} for lists, $\cup$ for sets,
$\uplus$ for multisets). These
operators can not be defined without using universal quantifiers
(or recursion) with the languages analyzed in this
paper~\cite{DPP00-nancy}.

\baselineskip 10pt \small

\subsection*{Acknowledgments}

The authors wish to thank Alberto Policriti, Ashish Tiwari, and
Silvia Monica for useful discussions on the topics of this paper.
The anonymous referee greatly helped us in improving the
presentation of the paper. This work is partially supported by
MIUR project \emph{Ragionamento su aggregati e numeri a supporto
della programmazione e relative verifiche}.\vspace*{-2ex}


\newpage
\appendix

\section{Proofs of Model Properties}\label{modellistiche}

We recall some technical definitions.
Given two $\Sigma$-structures ${\cal A}$ and ${\cal B}$, $ {\cal
B} = \langle B, (\cdot)^{\cal B}\rangle$ is a \emph{substructure}
of ${\cal A} = \langle A, (\cdot)^{\cal A}\rangle$ if $B \subseteq
A$ and for all $x \in B$ it holds that $(x)^{\cal A} = (x)^{\cal B}$.
Given two
$\Sigma$-structures ${\cal A}$ and ${\cal B}$, a function
$h:A\longrightarrow B$ is said to be an  \emph{homomorphism} from
$\cal A$ to $\cal B$ if: $(i)$ $\forall f\in {\cal F}, a_1,\dots,
a_n\in A \: (h(f^{\cal A}(a_1,\dots,a_n))=f^{\cal
B}(h(a_1),\dots,h(a_n))) $ and $(ii)$ $ \forall p\in \Pi,
a_1,\dots,a_m\in A \:(p^{\cal A}(a_1,\dots,a_m)\rightarrow p^{\cal
B}(h(a_1),\dots,h(a_m)))\,.$ $h$ is said to be an
\emph{isomorphism} if $f$ is bijective and in the property~$(ii)$
also the $\leftarrow$ implication holds. Given  two
$\Sigma$-structures $\cal A$ and $\cal B$\/, an \emph{embedding} of
$\cal A$ in $\cal B$ is an isomorphism from $\cal A$ to a
substructure of $\cal B$\/.

\begin{lemma}[\cite{CK73}]\label{logica}
Let $\cal A$ and $\cal B$ be two $\Sigma$-structures
and let $h$
be an embedding of $\cal A$ in $\cal B$\/.
If $\varphi$ is an open formula of ${\cal
L} = \langle \Sigma, {\cal V}\rangle $\/,
then for each valuation $\sigma$ on $A$
it holds that:
$${\cal A}\models \sigma(\varphi) \leftrightarrow {\cal
B}\models h(\sigma(\varphi))\,.$$
\end{lemma}

\begin{lemma}\label{versofacile}
$\cal MSET$ is a model of the theory \bag.
\end{lemma}

\begin{proof}
For each axioms/axiom schemata $(A)$ of the theory \bag\ we need
to prove that ${\cal MSET}$ models $(A)$ (briefly, ${\cal MSET}
\models (A)$). We give only the sketch of the proof.
\begin{description}

\item[$(K),(W)$\/:]
The fact that ${\cal MSET}$ is a model of $(K)$ and $(W)$ is a
consequence of the interpretation of the membership predicate in
$\cal MSET$ (cf.\ point (4) of Def.~\ref{defstruttura}).

\item[$(F^m_1)$\/:] This axiom holds in $\cal MSET$\/, since
$f(t_1,\dots,t_n)$ and $f(s_1,\dots,s_n)$ can be in the same class
in $\cal MSET$\/, only if for all $i=1,\dots, n$ it holds that
$t_i$ and $s_i$ belong to the same class.

\item[$(F_2)$\/:] It holds trivially, by definition of $\cal
MSET$\/, since terms beginning with different free symbols belong
to different classes.

\item[$(F_3),(F_3^m)$\/:]
The fact that ${\cal MSET} \models (F_3)$ and ${\cal MSET} \models
(F_3^m)$ holds in virtue of the finite size of each ground term;
it can be formally proved by induction on the complexity of the
terms.

\item[$(E_p^m)$\/:]
$\cal MSET$ is a model of $(E_p^m)$\/, since for any equational
theory $E$\/, $T({\cal F})/\equiv_E$ is a model of
$E$~\cite{Sie90}.

\item[$(E_k^m)$\/:]
$\cal MSET$ is a model of $(E_p^m)$\/, as seen in the previous
point, but it is also the \emph{initial} model, namely two terms
$s$ and $t$ are in the same class if and only if $(E_p^m)$ can
prove that $s=t$\/. This is exactly the meaning of the axiom
$(E_k^m)$.
\end{description}
\end{proof}

\begin{lemma}\label{versodifficile}
If $\cal M$ is a model of  \bag, then the function $ h:  T({\cal
F}_{MSet})/\equiv_{E_{MSet}}  \longrightarrow  { M}$\/, defined as
$h(\rappr{t})  = t^{\cal M}$ is an embedding of ${\cal MSET}$ in
${\cal M}$. \par\smallskip\par\noindent
\end{lemma}
\begin{proof}
We will prove the following facts:

\begin{enumerate}
\item The definition of $h(\rappr{t})$ does not depend on the
choice of the representative of the class;
\item $h$ is an homomorphism;
\item $h$ is injective;
\item if $h(\rappr{t})\in^{\cal M}h(\rappr{s})$\/,
then $\rappr{t}\in^{\cal MSET}\rappr{s}$\/.
\end{enumerate}

\noindent These facts imply the thesis.

\begin{enumerate}
\item If $t_1$ and $t_2$ are two terms such that
$\rapprmedio{t_1}=\rapprmedio{t_2}$\/, then by
definition $(E_p^m)\models t_1=t_2$\/.
Since ${\cal A}\models t_1=t_2$ holds in
every model ${\cal A}$ of
$(E_p^m)$\/, then in particular it holds in
$\cal M$\/, i.e.,
$t_1^{\cal M}=t_2^{\cal M}$\/.

\item We need to prove that:
\begin{enumerate}
\item for all $f\in {\cal F}_\bag$
and for all terms $t_1,\dots,t_n \in T({\cal F}_\bag)$
it holds that
$$h(f^{{\cal MSET}}(\rapprmedio{t_1},\dots,\rapprmedio{t_n})) =  f^{\cal
M}(h(t_1),\dots,h(t_n))$$ Now,
$$\begin{array}{rcll}
h(f^{{\cal MSET}}(\rapprmedio{t_1},\dots,\rapprmedio{t_n})) & = &
h(f(t_1,\dots,t_n)) & \mbox{By fact (1) above}\\
&=& (f(t_1,\dots,t_n))^{\cal M} & \mbox{By def. of
$h$}\\
&=& f^{{\cal M}}(t_1^{\cal
M},\dots,t_n^{\cal M}) & \mbox{By def.
of structure} \\
&=&  f^{{\cal M}}
(h(t_1),\dots,h(t_n))& \mbox{By def. of $h$}
\end{array}$$
\item for all terms $t$ and $s$,
if $\rappr{t}\in^{{\cal MSET}}\rappr{s}$\/, then
$h(\rappr{t})\in^{\cal M}h(\rappr{s})$. From $\rappr{t}\in^{{\cal
MSET}}\rappr{s}$, using fact 1. above, we have that there is a
term $s'$ in $\rappr{s}$ of the form $\mo t\:|\:r\mc$ and that
$h(\rappr{s})=s'^{\cal M}$\/. Hence, we have that
$h(\rappr{s})=\mo t^{\cal M}\,|\,r^{\cal M}\mc^{\cal M}$\/; $(W)$
ensures that $h(\rappr{t})=t^{\cal M}$ belongs to it.
\end{enumerate}

\item We prove,  by structural induction on $t_1$,
that if  $h(\rapprmedio{t_1})=h(\rapprmedio{t_2})$\/, then
$\rapprmedio{t_1}=\rapprmedio{t_2}$\/.

\Base Let $t_1$ be a constant $c$.
Since $\cal M$ is a model of axiom schema $(F_2)$\/, it can
not be
that $t_2 =  f(s_1,\dots,s_n)$\/, with $f$ different from $c$\/.
Hence, it must be that $t_2 = c$\/.

\Step Let $t_1$ be $f(s_1,\dots,s_n)$\/, with $f\not\equiv
\mf$\/.
It cannot be $t_2\equiv g(r_1,\dots,r_m)$\/, with
$g\not\equiv
f$\/, since ${\cal M}$ is a model of $(F_2)$\/.
So, it must be $t_2\equiv f(r_1,\dots,r_n)$\/, and, by $(F_{1})$,
$s_i^{{\cal M}}= r_i^{{\cal M}}$\/, for all $i\leq n$\/.
Using the inductive hypothesis
we have $\rapprmedio{t_1}=\rapprmedio{t_2}$\/.

Let $t_1$ be $\mo s_1,\dots, s_n \,|\, r\mc$\/, with $r$ not of
the form $\mo r_1\,|\,r_2\mc$\/. Since it cannot be that
$t_2$ is  $f(v_1,\dots, v_n)$ (from the previous case applied to
$t_2$), then it must be $t_2$ is $\mo u_1,\dots,  u_m\,|\,
v\mc$\/, for some $v$ not of the form $\mo v_1\,|\,v_2\mc$\/. Let
us assume, by contradiction, that $\rapprmedio{t_1}\neq \rapprmedio{t_2}$\/, and
$t_1^{\cal M}=t_2^{\cal M}$\/, while the thesis holds for all
terms of lower complexity. From $t_1^{\cal M}=t_2^{\cal M}$ we
obtain that the two terms have in $\cal M$ the same elements.
Since $\cal M$ is a model of $(W)$\/, the elements of $t_1^{\cal
M}$ are exactly $s_1^{\cal M},\dots, s_n^{\cal M}$ and the
elements of $t_2^{\cal M}$ are exactly $u_1^{\cal
M},\dots,u_m^{\cal M}$\/. So, by inductive hypothesis, there is a
bijection $b:\{1,\dots,n\}\longrightarrow \{1,\dots,m\}$ such
that $\rapprmedio{s_i}=\rapprmediocre{u_{b(i)}}\,\,$\/. This means that $m=n$ and that there
is a term $t_2'$ in $\rapprmedio{t_2}$ of the form $\mo s_1,\dots, s_m\,|\,
v\mc$\/. Applying $n$ times $(E_k^m)$\/, in all possible ways, we
obtain that $r^{\cal M}=v^{\cal M}$\/, hence by inductive
hypothesis $\rappr{r}=\rappr{v}$\/. From this fact, we conclude that
$\rapprmedio{t_2}=\rapprmedio{t_2'}=\rapprlungo{\mo s_1,\dots,s_n\,|\,r\mc}=\rapprmedio{t_1}$\/, which is in
contradiction with our assumption.

\item If $h(\rappr{t})\in^{\cal M}h(\rappr{s})$\/, then $t^{\cal M}\in^{\cal
M}s^{\cal M}$ and hence $(K)$ implies that $s$ must be a term of
the form $\mo t_1\,|\,t_2\mc$\/. By induction on $s$ using
$(W)$\/, we can prove that in particular $s$ must be a term of the
form $\mo t_1,\dots, t_i,\dots ,t_n\,|\,r\mc$\/, with $t_1^{\cal
M}=t^{\cal M}=h(\rappr{t})$\/. We have already proved that $h$ is
injective, hence it must be $t_1\in\rappr{t}$\/, and from this we
obtain $\rappr{t}\in^{\cal MSET}\rappr{s}$\/.
\end{enumerate}
\end{proof}

\begin{lemma}\label{convergente}
If $C$ is a constraint in pre-solved form and acyclic,
then $\sigma_C$ is
stabilizing.
\end{lemma}
\begin{proof}
We prove that $\sigma_C^*\equiv \sigma_C^{q-1}$\/, where $q$ is
the number of variables which occur in the right-hand side of
membership atoms.

The acyclicity condition ensures that there are no loops in the
graph ${\cal G}_{C}^{\in}$\/. Consider now the substitution
$\sigma_C$ and let $B$ be the set of the nodes of the graph that
belong to its domain (we identify variables and corresponding
nodes). Each application of $\sigma_C$ on the terms of its
codomain can be intuitively mimicked by a game that updates the
value of $B$ with the nodes corresponding to the variables
occurring in the terms $\sigma_C(B)$. These nodes can be computed
by collecting the nodes that can be reached by crossing an edge
from a node of $B$ (new variables $F_i,M_i$ are all different, and
they are not in the domain of $\sigma_C$, so we can forget them).
The process will terminate when either $B$ is empty or it contains
only variables that are not in the domain of $\sigma_C$. Since
${\cal G}_{C}^{\in}$ is acyclic, this process must terminate, and
since the longest path in the graph is shorter than $q$, it is
plain o see that $q-1$ is an upper bound to the number of
iterations.
\end{proof}

\begin{lemma}\label{listisolved}
Let $\Th$ be one of the theories \lst, \bag, \clist, ${\cal
A}_{\Th}$ the model (structure) which corresponds with $\Th$\/, and
$E_\Th$ the associated equational theory. Let $t, t'$ be two
terms and $C$ a solved form constraint over the language ${\cal
L}_\Th$, such that $FV(t)\cup FV(t')\subseteq FV(C)$. If ${\cal
A}_{\Th} \not\models{\forall}(t=t')$, then
$E_\Th\not\models{\forall}(\sigma_C^*(t)=\sigma_C^*(t'))$\/.
\end{lemma}
\begin{proof}
Let $R=\{X_1,\dots,X_n\}$ be the set of variables over which
$\sigma_C$ is defined.
By induction on the sum of the complexities of $t$ and $t'$ we prove
the following property that
implies the thesis of the lemma.
\begin{quote}
{\it
If there exists $\theta$ such that ${\cal A}_{\Th}\models \theta(t)\neq
\theta(t')$, then there exists $\theta'$ such that ${\cal A}_{\Th}\models
\theta'(\sigma_C^*(t)) \neq \theta'(\sigma_C^*(t'))$.}
\end{quote}

Let us consider the valuation $\theta''$ defined as:
$$\theta''(Y)=\left\{\begin{array}{ll}\theta(Y) & \mbox{if }Y\not\in R\\
                           \theta(X_i) & \mbox{if }Y\equiv M_{X_i}
                           \end{array}\right .$$
Observe that $\theta''$ is not defined over the variables $F_{X_1},\dots,
F_{X_n}$.

Let $m=\max\{size(\theta''(\sigma_C^*(t))),
size(\theta''(\sigma_C^*(t')))\}+1$. We can now define the
valuation $\theta'$ in the following way:
$$\theta'(Y)=
\left\{
\begin{array}{ll}
{[} \nil ]^{m*i} ({\mo} \nil \mc^{m*i}, {\clo} \nil \clc^{m*i}) &
      \mbox{if }Y\equiv F_{X_i}\\
\theta''(Y) & \mbox{otherwise }
\end{array}\right .$$
\noindent If $t = Y_1$ and $t'= Y_2$ are variables then:
\begin{itemize}
\item if $\sigma_C$ is not defined neither on $Y_1$ nor on $Y_2$, then
$\theta'(\sigma_C^*(Y_1))=\theta'(Y_1)=\theta(Y_1)\neq
\theta(Y_2)=\theta'(Y_2)=\theta'(\sigma_C^*(Y_2))$;
\item if $\sigma_C$ is defined on $Y_1$ and not on $Y_2$ (or
viceversa), then $\size(\theta'(\sigma_C^*(Y_1)))\geq \size(\theta'(F_{Y_1}))>
size(\theta'(Y_2))$;
\item if $\sigma_C$ is defined both on $Y_1$ and on $Y_2$, then:
\begin{description}
\item[\lst\ {\rm and} \clist:]
$\theta'(\sigma_C^*(Y_1))$
and $\theta'(\sigma_C^*(Y_2))$ differ on their first element.
\item[\bag:]
$\theta'(\sigma_C^*(Y_1))$
and $\theta'(\sigma_C^*(Y_2))$ differ on their elements $\theta'(F_{Y_1})$
and $\theta'(F_{Y_2})$.
\end{description}
\end{itemize}

\noindent If $t = Y$ is a variable and $t'$ is $f(t_1',\dots, t_h')$,
also when $f$ is of the form  $\lf,\lcf,\mf$, then:
\begin{itemize}
\item if $\sigma_C$ is not defined on $FV(t')\cup Y$, then we have
immediately the thesis since $\theta'(\sigma_C^*(Y))=\theta(Y)$
and $\theta'(\sigma_C^*(t'))=\theta(t')$;
\item if $\sigma_C$ is defined on $Y$, but not on $FV(t')$, then
we have the thesis since
$size(\theta'(\sigma_C^*(Y)))>size(\theta(t'))$;
\item if $\sigma_C$ is defined on at least one variable of $t'$
and not on $Y$, then as in the previous case we have the thesis;
\item if $\sigma_C$ is defined on $Y$ and on at least one of the
variables of $t'$, then:
\begin{description}
\item[$\lst$ {\rm and} $\clist$:] it can never be the case that the first
element of $\theta'(\sigma_C^*(Y))$---i.e. $\theta'(F_Y)$---is
equal to the first element of $\theta'(\sigma_C^*(t'))$\/; this
follows from the conditions we have imposed on all the
$\theta'(F_{X_i})$.
\item[$\bag$:] two cases are possible:
$\theta'(F_Y)$ is not an element of $\theta'(\sigma_C^*(t'))$,
from which we have the thesis.  \\
$\theta'(F_Y)$ is an element of $\theta'(\sigma_C^*(t'))$:
this means that ${\sf tail}(t')=Y$, hence the thesis follows.
\end{description}
\end{itemize}

\noindent If $t$ is  $f(t_1,\dots,t_h)$ and $t'= g(t_1',\dots,t_k')$,
with $f$ different from $g$, then it is trivial.

\noindent If $t$ is  $f(t_1,\dots,t_h)$ and $t'$ is $f(t_1',\dots,t_h')$,
with $f$ different from $ \lf,\lcf,\mf$, then by inductive hypothesis we have the
thesis.

\noindent If $t$ is  $[ t_1\,|\,t_2]$ and
$t'$ is $[t_1'\,|\,t_2']$, then from ${\cal LIST}\models\theta(t)\neq
\theta(t')$ we have that it must be ${\cal LIST}\models\theta(t_1)\neq
\theta(t_1')$ or ${\cal LIST}\models\theta(t_2)\neq
\theta(t_2')$, hence, in both cases, we obtain the thesis by inductive
hypothesis.

\noindent If $t$ is $\clo t_1\,|\,t_2\clc$ and $t'$ is
$\clo t_1'\,|\,t_2'\clc$, then from ${\cal CLIST}\models\theta(t)\neq
\theta(t')$ we have that it must be ${\cal CLIST}\models\theta(t_1)\neq
\theta(t_1')$ or ${\cal CLIST}\models\theta(t_2)\neq
\theta(t_2')\wedge \theta(t_2)\neq
\clo\theta(t_1')\,|\,\theta(t_2')\clc\wedge \theta(t_2')\neq
\clo\theta(t_1)\,|\,\theta(t_2)\clc$, hence:
\begin{itemize}
\item in the first case we obtain the thesis by inductive
hypothesis on $t_1$ and $t_1'$.
\item in the second case by inductive hypothesis on $t_2$ and
$t_2'$, on $t_2$ and $\clo t_1'\,|\,t_2'\clc$, on $t_2'$ and $\clo
t_1\,|\,t_2\clc$, we obtain that ${\cal CLIST}\models \theta'(\sigma_C^*(t_2))
\neq \theta'(\sigma_C^*(t_2'))$ and ${\cal CLIST}\models
\theta'(\sigma_C^*(t_2))
\neq \theta'(\sigma_C^*(\clo t_1'\,|\,t_2'\clc))$ and ${\cal CLIST}\models
\theta'(\sigma_C^*(t_2'))
\neq \theta'(\sigma_C^*(\clo t_1\,|\,t_2\clc))$, which implies our
thesis.
\end{itemize}

\noindent If $t$ is  $\mo t_1\,|\,t_2\mc$ and $t'$ is
$\mo t_1'\,|\,t_2'\mc$, then:
\begin{itemize}
\item if ${\sf tail}(t_2)$ and ${\sf tail}(t_2')$ are the same
variable, the we obtain the thesis by inductive hypothesis on
${\sf untail}(\mo
t_1\,|\,t_2\mc)$ and ${\sf untail}(\mo t_1'\,|\,t_2'\mc)$;
\item if ${\sf tail}(t_2)=Y$ and ${\sf tail}(t_2')=Y'$ are not the
same variable and $\sigma_C$ is not defined on $Y$ or on $Y'$,
then $\theta'(F_Y)$ or $\theta'(F_{Y'}$ is not an element of both
$\theta'(\sigma_C^*(t))$ and $\theta'(\sigma_C^*(t'))$;

\item if ${\sf tail}(t_2)=Y$ and ${\sf tail}(t_2')=Y'$ are not the
same variable and $\sigma_C$ is not defined on $Y$ and on $Y'$,
then we can restrict ourselves to the case in which there is an
element $s$ of $\theta(t)$ which is not an element of
$\theta(t')$ (in the general case we would have to consider that
there exists $s$ such that there are $m$
occurrences of $s$ in $\theta(t)$ and $n$ occurrences in
$\theta(t')$ with $m\neq n$):
\begin{itemize}
\item if $s$ is an element of $\theta(Y)$, then, from the fact that
$\sigma_C$ is not defined on $Y$, we have the thesis, since it cannot be
the case that one of the elements of ${\sf untail}(t')$ becomes equal to
$\theta(s)$ (the new elements have a size which is greater);
\item if $s$ is an element of ${\sf
untail}(t)$, then we have $t = \mo u_1,\dots,
u_h,\dots,u_m\,|\,Y\mc$ and $s = \theta(u_h)$, hence, from the
inductive hypothesis, we have that $\theta'(\sigma_C^*(u_h))$ is
still different from all elements of $\theta'(\sigma_C^*({\sf
untail}(t')))$, and it is immediate that it is different from all
the elements of $\theta'(Y')$, hence $\theta'(\sigma_C^*(u_h))$ is
an element of $\theta'(\sigma_C^*(t))$ which is not in
$\theta'(\sigma_C^*(t'))$.
\end{itemize}
\end{itemize}
\end{proof}

\begin{lemma}\label{uffa}
Let $\Th$ be one of the theories \lst, \clist, \bag\ and \set,
and $C$ a constraint in pre-solved form over the language of
$\Th$. If ${\sf is\_solved_\Th(C)}$ returns $\false$, then $C$ is
not satisfiable in the model ${\cal A}_{\Th}$
which corresponds with $\Th$.
\end{lemma}
\begin{proof}
If ${\sf is\_solved_\Th(C)}$ returns $\false$ because ${\cal
G}_{C}^{\in}$ has a cycle then the result is trivial, since all
aggregates in $\cal A$ are well-founded. Otherwise:
\begin{description}
\item[{\rm For} \lst, \bag, \clist: ] From Lemma \ref{listisolved} we know
that $\Th\models\forall(\sigma_C^*(t)=\sigma_C^*(t'))$
implies ${\Th}\models \forall(t=t')$, hence, since $t\in
X$ and $t'\not\in X$ are in $C$, $C$ is not satisfiable in the
model $\cal A$ which corresponds with $\Th$.
\item[{\rm For} \set: ] Let
$\sigma_C^*\equiv [X_1/\{ F_1, p_1^1,\dots, p_1^{k_1}\,|\,M_1\},\dots,
X_q/\{F_q,p_q^{1},\dots,p_q^{k_q}\,|\,M_q\}]$, we have that
if ${\cal SET}\models C\gamma$, then ${\cal SET}\models
(C\sigma_C^*)\gamma'$, where $\gamma'$ is defined as follows
$$\gamma'(Y)=\left\{\begin{array}{ll}
\gamma(X_i) & \mbox{ if } Y\equiv M_i\\
p_i^1 & \mbox{ if } Y\equiv F_i\\
\gamma(Y) & \mbox{ otherwise}
\end{array}\right.  $$
Hence, if ${\sf is\_solved_\set}$ returns $\false$ this means that
$C\sigma_C^*$ is not satisfiable in $\cal SET$, which implies
that $C$ is not satisfiable in ${\cal SET}$.
\end{description}
\end{proof}

\section{Termination Proofs (Theorem \ref{termina-lists})} \label{termination}

\subsection*{Termination of ${\sf SAT}_{\lst}$}
Using the same measure as for ${\sf SAT}_{\bag}$ termination follows.\qed

\subsection*{Termination of ${\sf SAT}_{\clist}$}
Finding a global decreasing
measure implies that this measure is decreased by each rule of each
algorithm involved. The measure developed in~\cite{DPR98-fuin} for proving
termination of {\sf Unify\_clists} is rather complex. This is due to the
fact that new variables are (apparently) freely introduced in the
constraint by this procedure. Instead of extending such complex measure to
the general case, we use here a different approach for proving termination.
The proof is based:
\begin{itemize}
\item on the fact that each single rewriting procedure
terminates
(for {\sf Unify\_clists} it follows from~\cite{DPR98-fuin};
for the other three procedures the result is trivial) and
\item on the fact that it is possible to find a bound on
the number of possible {\sf repeat} cycles.
\end{itemize}

The remaining part of the proof is devoted to find this
bound.
First of all observe that:
\begin{itemize}
\item After the execution of {\sf in-CList} there are only
membership atoms of the form $t \in X$ with $X \notin
\vars(t)$\/. New equations can be introduced.
\item After the execution of {\sf in-CList} there are only
not-membership literals of the form $t \notin X$ with $X \notin
\vars(t)$\/. New disequality constraints can be introduced. Membership atoms
are not introduced.
\item After the execution of {\sf neq-CList} there are
only disequality constraints of the form $X \neq t$ with $X \notin
\vars(t)$\/. New equations can be introduced. $\in$ and
$\notin$-constraints are not introduced.
\item {\sf Unify\_clists} eliminates all equality
constraints producing a substitution. This substitution, when
applied to membership, not-membership, and disequality literals in
pre-solved form  can force a new execution of the procedures {\sf
in-CList}, {\sf nin-CList}, and {\sf neq-CList}. However, new
executions of {\sf Unify\_clists} are possible only if {\sf
in-CList} and {\sf neq-CList} introduce new equations. In the
following we will find a bound on the number of possible new
equations inserted.
\end{itemize}

Let us analyze membership constraints. Each membership atom of
the form $t \in \clo s' \,|\, s'' \clc$ is rewritten to \false\
or to $t = s' \vee t \in s''$\/. This means that in each
non-deterministic branch of the rewriting process \emph{at most}
one equation is introduced for each initial membership atom. Thus,
if $k$ is the number of membership atoms in $C$ at the beginning
of the computation, at most $k$ equality atoms (that can fire
{\sf Unify\_clists}) can be introduced. If we prove termination
with $k = 0$ then full termination easily follows, since it is the
same as considering $k$ successive (terminating) executions.

Let us consider the procedure {\sf neq-CList}. Action $(7.2)$ can
replace a disequality constraint of the form: $X \neq \clo
t_1,\dots,t_n \,|\, X \clc$ with the following equations,
identifying a substitution:
\begin{eqnarray}
\label{Xnil} X & = & \nil\\
\label{Xcli}
X & = & \clo N_1\,|\,N_2 \clc\,\mbox{ with $N_1,N_2$ new
variables.}
\end{eqnarray}

Let us analyze the various cases in which substitutions of
this form have
some effects on the constraint.
\begin{itemize}
\item there is $t \in X$ in $C$\/. This is not possible by
hypothesis,
  since $k=0$\/.
\item $t \notin X$ or $X \neq t$ and we know that $X$ does not
 occur in $t$\/. This implies a finite number of executions
  of rules of
  {\sf nin-CList} or {\sf neq-CList}. Since $X$ is not in $t$ and
  the variables $N_1$ and $N_2$ are newly introduced, it is impossible
  to generate a situation firing rule $(7.2)$\/.
\item Assume there are more than one equation introduced for
the same variable $X$\/.
\begin{itemize}
\item If they are all of the form (\ref{Xnil}), then
  {\sf  Unify\_clists} will apply the substitution and
  remove the redundant equations.

\item If they are all of the form (\ref{Xcli}), then
  {\sf  Unify\_clists} will perform a unification process between
  these new equations. The particular form of the equations allows us
  to see that the effect is to introduce new equations of the form
  $N_1 = N'_1$ between all the new variables used as elements
  and equations of the form $N_2 = N'_2$ or
  $N_2 = \clo N'_1 \,|\, N'_2 \clc$ between the new
variables used as rests.
  The situation is similar to that in which a unique
substitution is computed.

\item  If there are both equations of the form
 (\ref{Xnil}) and of the form  (\ref{Xcli}), then
 a failing (thus, terminating) situation will be detected by
{\sf Unify\_clists}.\qed

\end{itemize}
\end{itemize}

\subsection*{Termination of ${\sf SAT}_{\set}$}

Finding a global decreasing measure
implies that this measure is decreased by each rule of each algorithm
involved. This  is rather complex since it must subsume the measure
developed in~\cite{DPR98-fuin} for proving termination of {\sf
Unify\_sets}. Thus, instead of extending such complex measure, we use here
a different approach for proving termination. The proof is based:
\begin{itemize}
\item on the fact that each single rewriting procedure
terminates
(for {\sf Unify\_sets} it follows from~\cite{DPR98-fuin};
for the other three procedures the result is trivial) and
\item on the fact that it is possible to control the number
of new
calls to unification.
\end{itemize}
In order to simplify the proof we assume a strategy for handling the
non-determinism. The strategy will be pointed out during the discussion.

As observed in the proof of ${\sf SAT}_{\clist}$\/, if $k$ is the
number of membership atoms in $C$ at the beginning of the
computation, at most $k$ equality atoms (that can fire {\sf
Unify\_sets}) can be introduced. For this reason, we can safely
forget this kind of constraints from the whole reasoning.

The only problem for termination is given by rules $(6a)$ and
$(6b)$  of {\sf neq-CList}. As a strategy, we can unfold the
application of this rules (actually, adding a bit of determinism
to the whole procedure). This means that rule $(6a)$ (for $(6b)$
the situation is symmetrical) is as follows:
   assume that $\{ t_1  \,|\, s_1 \}$ is $\{
v_1,\dots,v_m\,|\,h\}$ and
   $\{ t_2 \,|\, s_2 \}$ is $\{w_1,\dots,w_n\,|\,k\}$\/,
with $h,k$
   variables or terms of the form $f(\dots),g(\dots)$\/, $f$
and $g$
   different from $\sef$\/.
   The global effect of the subcomputation is that of
returning a
   constraint of the form ($1 \leq i \leq m$):
   \begin{eqnarray}
   \label{primoset}
   N = v_i, v_i \neq w_1, \dots, v_i \neq w_n, v_i \notin k
\end{eqnarray}
   or one constraint of the form
   \begin{eqnarray}
   \label{secondoset}
   h = \{ N \,|\, N'\}, N \neq w_1,\dots,N \neq w_n, N
\notin k
   \end{eqnarray}
  if  $h$ is a variable. Notice that the application of this
  substitution is a sort of application of rule $(4)$ of the
  procedure {\sf in-Set}.

In the following discussion let us assume that termination by
failure do not occur (but, in this case, termination follows
trivially). Suppose to have already executed the first cycle of
the {\sf repeat} loop. Local termination ensures that this can be
done in finite time. In the constraint there are no equations,
while there can be negated membership and disequality literals not
necessarily in pre-solved form.

Let us execute procedure {\sf nin-Set}. No equations are
introduced. In the constraint there are not-membership literals in
pre-solved form and disequality constraints not necessarily in pre-solved
form.

Let us execute the procedure {\sf neq-Set}.
We adopt a  weak strategy to face the
non-determinism:
delay the constraints that fire action $(6)$ as much as
possible.
This means that after a finite time the constraint is
composed
by a number of constraints of the form $X \neq t$ or
$t \notin X$ with $X\notin \vars(t)$ plus a (possibly empty)
constraint $\tilde C$ of constraints all firing action $(6)$
of
{\sf neq-Set}. Pick one constraint $c$ from $\tilde C$
and consider the possible non-deterministic executions.

\begin{itemize}
\item Assume that the situation of
   case (\ref{secondoset}) above does not occur in a
   non-deterministic branch. Then (see case (\ref{primoset}))
   the constraint  $c$ is replaced in $ C$ by
   a number of constraints $v_i \neq w_j$ of fewer size. If
   they do not fire
   action $(6)$ they can be directly
   processed to reach a pre-solved form.
   Otherwise, they are inserted in $\tilde C$\/, but since
   they are of fewer size, if the situation of
   case (\ref{secondoset}) never occur, this again implies
   termination.
\item Assume now that the situation of
   case (\ref{secondoset})  occurs when processing
   the constraint  $c$\/. Constraints
   $$N \neq w_1,\dots,N \neq w_n, N \notin k$$
   are introduced. Constraints in pre-solved form
   of the form above, with $N$ a variable introduced as element
   of a set by action $(6)$\/, are said \emph{passive constraints}.
   Variables $N$ of this form are inserted in the constraint
   only by this step. We will see that passive disequality constraints
   remain in pre-solved form forever while negated membership
   passive literals have a controlled growth.

   Assume to apply immediately the
   substitution $h / \{ N \,|\, N' \}$\/. Its effect can be
   the following, according to the position of $h$ in a
constraint:
   \begin{itemize}
   \item $X \neq t[h]$ or $t[h] \notin X$\/: the terms gets
      changed
      but the constraints remain in pre-solved form.
   \item $s[h] \neq t $ or $s \neq t[h]$
        or $s[h] \neq t[h]$\/: the terms change
      but the constraints remain in $\tilde C$ to be
      processed
          later.
   \item  $t  \notin h$ is transformed to
       $t  \notin \{ N \,|\, N' \}$\/.
           One step of {\sf nin-Set} is applied to obtain:
           $t \neq N \wedge t  \notin  N'$\/.
           The first constraint is immediately transformed into
           $N \neq t$ (a passive constraint) while the second is in
       pre-solved form. Observe that if  $t  \notin h$  is
passive  (i.e., $t$ is a variable of type $N$\/), then only
passive constraints are introduced.

   \item $h \neq t$ is transformed to $\{ N \,|\, N' \} \neq
       t$\/. Observe that $h \neq t$ can not be a passive constraint
           since $h$ is a `rest' variable while the variables
           of passive constraints are `element' variables, like $N$ here.
           A constraint in pre-solved form is no longer in pre-solved
           form. Let us apply the rewriting rules to it.
           It is immediately rewritten to \true\ (e.g., when $t$ is
           $f(\cdots)$\/, $f \neq \sef$\/) or it becomes in pre-solved form
           (when $t$ is a variable) or action (6) can be applied.

           Both in cases (\ref{primoset}) and in the case
           (\ref{secondoset})
           we introduce a number of passive constraints and, in the
           last case, a substitution $N' / \{ N_1 \,|\, N_1' \}$
           is applied. Notice that the global effect on the system it
           the fact that in the other constraints the original variable
           $h$ is replaced by $\{N,N_1\,|\,N'_1\}$\/.
           This means that this situation can be performed at most once
           per each occurrence of $h$\/. And, the reasoning starting
           from substitutions of the form
           $\{N,N_1,\dots,N_{\ell}\,|\,N'_{\ell}\}$
           is the same as that done here for  $N' / \{ N_1 \,|\,N_1' \}$\/.
           At the end of the process,
           the number of constraints in $\tilde C$ is decreased and
           we have only introduced pre-solved form and passive constraints.\qed
   \end{itemize}
\end{itemize}

\end{document}